\documentclass{aastex62}
\turnoffedit
\pdfoutput=1

\usepackage{multirow}

\received{April 29, 2019}
\revised{May 28, 2019}
\revised{July 9, 2019}
\accepted{July 9, 2019}
\submitjournal{AJ}

\shorttitle{Venus' long-term 365-nm albedo variations}
\shortauthors{Lee et al.}

\begin{document}

\title{Long-term variations of Venus' 365-nm albedo observed by Venus Express, Akatsuki, MESSENGER, and Hubble Space Telescope}

\correspondingauthor{Yeon Joo Lee}
\email{yjleeinjapan@gmail.com}

\author[0000-0002-4571-0669]{Yeon Joo Lee}
\affil{Graduate School of Frontier Sciences, The University of Tokyo, Kashiwa, Japan}

\author{Kandis-Lea Jessup}
\affiliation{Southwest Research Institute, Boulder, USA}

\author{Santiago Perez-Hoyos}
\affiliation{Departamento F\'isica Aplicada I, Escuela de Ingenieria, Universidad del Pa\'is Vasco UPV/EHU, Bilbao, Spain}

\author{Dmitrij V. Titov}
\affiliation{ESTEC/ESA, Noordwijk, Netherlands}

\author{Sebastien Lebonnois}
\affiliation{LMD/IPSL, CNRS, Paris, France}

\author{Javier Peralta}
\affiliation{Institute of Space and Astronautical Science (ISAS/JAXA), Sagamihara, Japan}

\author{Takeshi Horinouchi}
\affiliation{Faculty of Environmental Earth Science, Hokkaido University, Sapporo, Japan}

\author{Takeshi Imamura}
\affil{Graduate School of Frontier Sciences, The University of Tokyo, Kashiwa, Japan}

\author{Sanjay Limaye}
\affiliation{Space Science and Engineering Center, University of Wisconsin, Madison, USA}

\author{Emmanuel Marcq}
\affiliation{LATMOS/IPSL, UVSQ Universit\'e Paris-Saclay, Sorbonne Universit\'e, CNRS, Guyancourt, France}

\author{Masahiro Takagi}
\affiliation{Department of Astrophysics and Atmospheric 4 Science, Faculty of Science, Kyoto Sangyo University, Kyoto, Japan}

\author{Atsushi Yamazaki}
\affiliation{Institute of Space and Astronautical Science (ISAS/JAXA), Sagamihara, Japan}
\affiliation{Department of Earth and Planetary Science, Graduate School of Science, The University of Tokyo, Tokyo, Japan}

\author{Manabu Yamada}
\affiliation{Planetary Exploration Research Center (PERC), Chiba, Institute of Technology, Narashino, Japan}

\author{Shigeto Watanabe}
\affiliation{Hokkaido Information University, Ebetsu, Japan}

\author{Shin-ya Murakami}
\affiliation{Institute of Space and Astronautical Science (ISAS/JAXA), Sagamihara, Japan}

\author{Kazunori Ogohara}
\affiliation{University of Shiga Prefecture, Hikone, Japan}

\author{William M. McClintock}
\affiliation{Laboratory for Atmospheric and Space Physics, Boulder, USA}

\author{Gregory Holsclaw}
\affiliation{Laboratory for Atmospheric and Space Physics, Boulder, USA}

\author{Anthony Roman}
\affiliation{Space Telescope Science Institute, Baltimore, USA}

\begin{abstract}
\added{An unknown absorber near the cloud top level of Venus generates a broad absorption feature from the ultraviolet (UV) to visible, peaking around 360~nm, and therefore plays a critical role in the solar energy absorption. We present a quantitative study on the variability of the cloud albedo at 365 nm and its impact on Venus’ solar heating rates based on an analysis of Venus Express and Akatsuki's UV images, and Hubble Space Telescope and MESSENGER's UV spectral data; in this analysis the calibration correction factor of the UV images of Venus Express (VMC) is updated relative to the Hubble and MESSENGER albedo measurements. Our results indicate that the 365-nm albedo varied by a factor of 2 from 2006 to 2017 over the entire planet, producing a 25--40\% change in the low latitude solar heating rate according to our radiative transfer calculations. Thus, the cloud top level atmosphere should have experienced considerable solar heating variations over this period. Our global circulation model calculations show that this variable solar heating rate may explain the observed variations of zonal wind from 2006 to 2017. Overlaps in the timescale of the long-term UV albedo and the solar activity variations make it plausible that solar extreme UV intensity and cosmic-ray variations influenced the observed albedo trends. The albedo variations might also be linked with temporal variations of the upper cloud SO$_2$ gas abundance, which affects the H$_2$SO$_4$-H$_2$O aerosol formation.}
\deleted{The ultraviolet (UV) to visible spectrum of Venus has a broad absorption feature which peaks at 360~nm due to the presence of an unknown absorber near the cloud top level (around 70~km). While the chemical composition of the unknown absorber has been a 4-decade old mystery, it is known that about half of the solar energy is deposited in the atmosphere due to this absorber. Thus, the unknown absorber plays a critical role in the atmospheric energy balance. Here we present a quantitative study on the variability of the cloud albedo at 365~nm and its impact on Venus’ solar heating rates based on an analysis of Venus Express and Akatsuki's UV images, and Hubble Space Telescope and MESSENGER's UV spectral data. These results show that the 365-nm albedo varied by a factor of 2 from 2006 to 2017 over the planet. Our radiative transfer calculations show that the observed 365-nm albedo variation can produce a 25--40\% change in the low latitude solar heating rate, and therefore that the cloud top level atmosphere should have experienced considerable solar heating variations over this period. We suggest that this variable solar heating may explain the reported Venus’ zonal wind variations from 2006 to 2017, and show the plausible process through global circulation model calculations. Further studies are required to understand possible climate change on the current Venus in response to the inferred solar heating variations.}
\end{abstract}


\section{Introduction} \label{sec:intro}
The Solar radiance is the principal energy source for the atmosphere of terrestrial planets, such as Earth, Venus, and Mars. Inhomogeneous solar radiance absorption drives atmospheric motions, from small scale convection to large scale global circulation. These motions distribute \replaced{excessive}{excess} energy, and transport mass and momentum in the atmosphere. Temporal variation of absorbed solar radiance is therefore an important indication of possible changes in the atmosphere. The long-term monitoring of solar energy absorption is particularly useful in radiative energy balance calculations as a major input energy into a planetary system.

On the atmosphere of Venus, the maximum solar energy deposition occurs in the upper cloud layer (60--70~km) rather than at the surface as in the case of the Earth \citep{Crisp86, Titov12}. The maximum solar energy absorption in the clouds is due to an unidentified absorber, hereafter ``unknown absorber'', which has been an unsolved question in Venus research regarding its identity. Venus’ global scale clouds and upper haze are mainly composed of H$_2$SO$_4$--H$_2$O \citep{Titov12, Allen64, Mills07} which has a small imaginary refractive index ($n_i=[1-9]\times10^{-8}$) in the UV to visible range \citep{Palmer75, Hummel88}. As a result, the H$_2$SO$_4$--H$_2$O clouds and haze absorb almost none of the solar radiance in this spectral range, but are effective scatterers, making a strong contribution to the total solar radiance scattered back to space (which is $\sim$75\% of the incident flux) \citep{Titov12}. UV images of Venus, however, show distinct \deleted{UV} patterns caused by the unknown absorber. The absorption spectrum produced by the unknown absorber is observed to reach maximum absorption levels between 340 and 380~nm, and then decrease smoothly with increasing wavelength from 380~nm through the visible range \citep{Barker75,Perezhoyos18}. Some studies indicate that an absorption tail exists at wavelengths shortward of 340~nm \citep{Perezhoyos18}, and in the 170--320~nm range \citep{Marcq11}. According to data from descent probes the unknown absorber may be located in the upper clouds \citep{Tomasko80a, Esposito80}, and absorbs about half of the solar radiance deposited at the cloud top level, accounting for $\sim$3~K/Earth day of the global mean solar heating around 65 km altitude, when the total global mean solar heating is $\sim$6~K/day \citep{Crisp86}. Many \deleted{unknown absorber} candidates have been proposed \added{for the unknown absorber,} including OSSO, S$_2$O, S$_x$, FeCl$_3$, and iron-bearing microorganism \citep{Mills07, Frandsen16, Krasnopolsky17, Perezhoyos18, Limaye18}. However, none of these species satisfy both the spectral features produced by the unknown absorber, and the lifetime and simulated vertical profile required to fit the observations \citep{Krasnopolsky18, Perezhoyos18}.

The UV patterns reveal clear temporal and spatial variations; for example, the well known strong zonal winds, or `super-rotation', that rotate around the globe in 4--5 days \citep{Barker75, Rossow90, Kouyama13}, and transport the unknown absorber horizontally. In addition to this background wind, Venus' infamous `Y-feature' is explained with atmospheric waves, resulting in a short-term periodicity of 4 to 5 days \citep{Boyer61, DelGenio82, DelGenio90, Peralta15, Kouyama15, Imai16}. In the mean time, short\deleted{-term} and long-term variations are also reported \citep{Khatuntsev13,Marcq13,Kouyama13, Lee15a}, which are tracked by the distribution or abundance of the unknown absorber. Temporal variations of latitudinal 365-nm contrasts are closely linked with the SO$_2$ gas abundance above the cloud top level \citep{Lee15a}, suggesting influences of chemical processes on sulfuric acid cloud aerosol formations in the UV contrast \citep{Esposito82,Parkinson15}. Since \citet{Marcq13} reported a general decline in the SO$_2$ gas abundance in the same periods where a decline in the long-term 365~nm cloud top albedo was observed, \citet{Lee15a} proposed that the rate of H$_2$SO$_4$ production, dependent on SO$_2$ photolysis, may be the principal mechanism supporting the observed long-term 365~nm albedo variation trends. However, definitive claims regarding these relationships were not made by \citet{Lee15a} due to uncertainties at that time regarding the impact of the instrument degradation on the retrieved 365~nm cloud top albedo \citep{Shalygina15}.

In this study, we report that the long-term variations of the UV reflectivity are a real \replaced{phenomena}{phenomenon} through a comparison of four space-based instruments\added{: imagers on board Venus Express and Akatsuki, and spectrometers on board MESSENGER and Hubble Space Telescope (Section~\ref{sec:data})}. We\deleted{Using the data described in Section~\ref{sec:data}, we} carefully checked the same phase angle disk-resolved data \replaced{(Sections~\ref{subsec:photo-corr} and \ref{subsec:disk-resolved-results})}{(Section~\ref{subsec:photo-corr})}\added{, and update the calibration correction factor of the UV data of Venus Express through a cross comparison with the UV spectra of MESSENGER and Hubble Space Telescope (Section~\ref{subsubsec:CrossComp_Disk-resA}). We evaluate the updated UV image data of Venus Express by a comparison with UV images of Akatsuki, showing a successful performance. We find common long-term variations in both of the disk-resolved albedo and the whole-disk albedo (Sections~\ref{subsec:disk-resolved-results} and \ref{subsec:whole-disk-results})}. \deleted{We also confirm the same trend in whole-disk (disk-integrated) albedo (Sections~\ref{subsec:whole-diskA} and \ref{subsec:whole-disk-results}).} The results are employed in our radiative transfer model calculations to understand possible solar heating variations (Sections~\ref{subsec:RTM} and \ref{subsec:RTMresults}). We discuss the significance of these results on the relationship with atmospheric winds at the cloud top level, \added{including Venus global circulation model calculations, }and possible reasons for the observed 365~nm albedo variations in Section~\ref{sec:discussion}.

\section{Data} \label{sec:data}
Our 365-nm reflectivity analysis covers a decade of period, from 2006 to 2017, with only a one-year gap in 2015. The data were acquired from four instruments. The longest period of monitoring, 2006--2014, was \replaced{done}{covered} with \added{the} Venus Monitoring Camera (VMC) on board Venus Express \citep{Markiewicz07a}. Two sets of 365-nm images were taken with \added{the} UV Imager (UVI) on board Akatsuki; one set of images was acquired from far distance in 2011 after the first Venus orbit insertion failure of Akatsuki \citep{Nakamura14}, and the other set of images was taken from 2015 December to 2017 May, after the successful orbit insertion \citep{Nakamura16}.
Regular star observations of UVI have been conducted from 2010 for the calibration purpose. This revealed steady sensitivity over time \citep{Yamazaki18}, and the mean error range in 2010--2017 is 18\%. Near-UV spectra were taken with \added{the} Mercury Atmospheric and Surface Composition Spectrometer (MASCS) on board MErcury Surface, Space ENvironment, GEochemistry, and Ranging (MESSENGER) during its Venus flyby on 2007 June 5, using the VIS-VIRS channel (320--950~nm, spectral resolution 4.7~nm) \citep{Perezhoyos18}. An error of 5-10\%  is estimated at 340--390~nm from star observations during MESSENGER's cruise phase \citep{Holsclaw10}. \replaced{Another}{Other} near-UV Venus spectra were taken with \added{the} Space Telescope Imaging Spectrograph (STIS) of Hubble Space Telescope with the G430L grating mode (290--570~nm, spectral resolution 0.54~nm) on 2011 January 2 \citep{Jessup15}. 5\% error is estimated for STIS measurements from based on regular standard star observations \edit2{\citep{Jessup19}}.

MASCS, STIS, and UVI data overlap observations with VMC data in 2007 and 2011. This is important for radiometric comparisons, as all of three performed star observations, and retrieved radiometric uncertainties independently. VMC data is highly uncertain in terms of radiometric calibration, as its star observations revealed 82\% error \citep{Titov12, Shalygina15}. Cross calibration between VMC and \replaced{a}{the} Visible and Infrared Thermal Imaging Spectrometer (VIRTIS) on board Venus Express was conducted, using simultaneous overlapped spectral range observations between VIRTIS and VMC \citep{Titov12, Shalygina15}. However, the absolute calibration of VIRTIS was not done at the time of these publications, resulting \replaced{a question on}{in questions about} this cross calibration \citep{Lee15a}. Comparisons in our study show that this VMC and VIRTIS cross calibration factor and the retrieved sensitivity degradation ratio of the 365-nm filter results in \replaced{a too large difference}{too large a difference} in VMC data with respect to MASCS, STIS and UVI (Section~\ref{subsubsec:CrossComp_Disk-resA}). That will be discussed in detail (Section~\ref{sec:methods}).

VMC images were manually selected in order to filter out those exhibiting artifacts, which could not be successfully corrected with additional flat field corrections \citep{Titov12}. In addition, we selected VMC and UVI images having the dayside of Venus completely within the field of view. There are two UVI flat fields of the CCD matrix, and this study used the one measured on the ground before the launch of Akatsuki \citep{Yamazaki18}. The same data were used for the star flux calibration \citep{Yamazaki18}, and the mean calibration correction factor ($\beta$) is 1.525 in 2010-2017. We multiplied this calibration correction factor $\beta$ \replaced{to}{by} the measured radiance of UVI. The other UVI flat field data were made with the on-board diffuser, and the public data in DARTS and PDS are generated using the on-board diffuser flat field.

\begin{figure}
\plotone{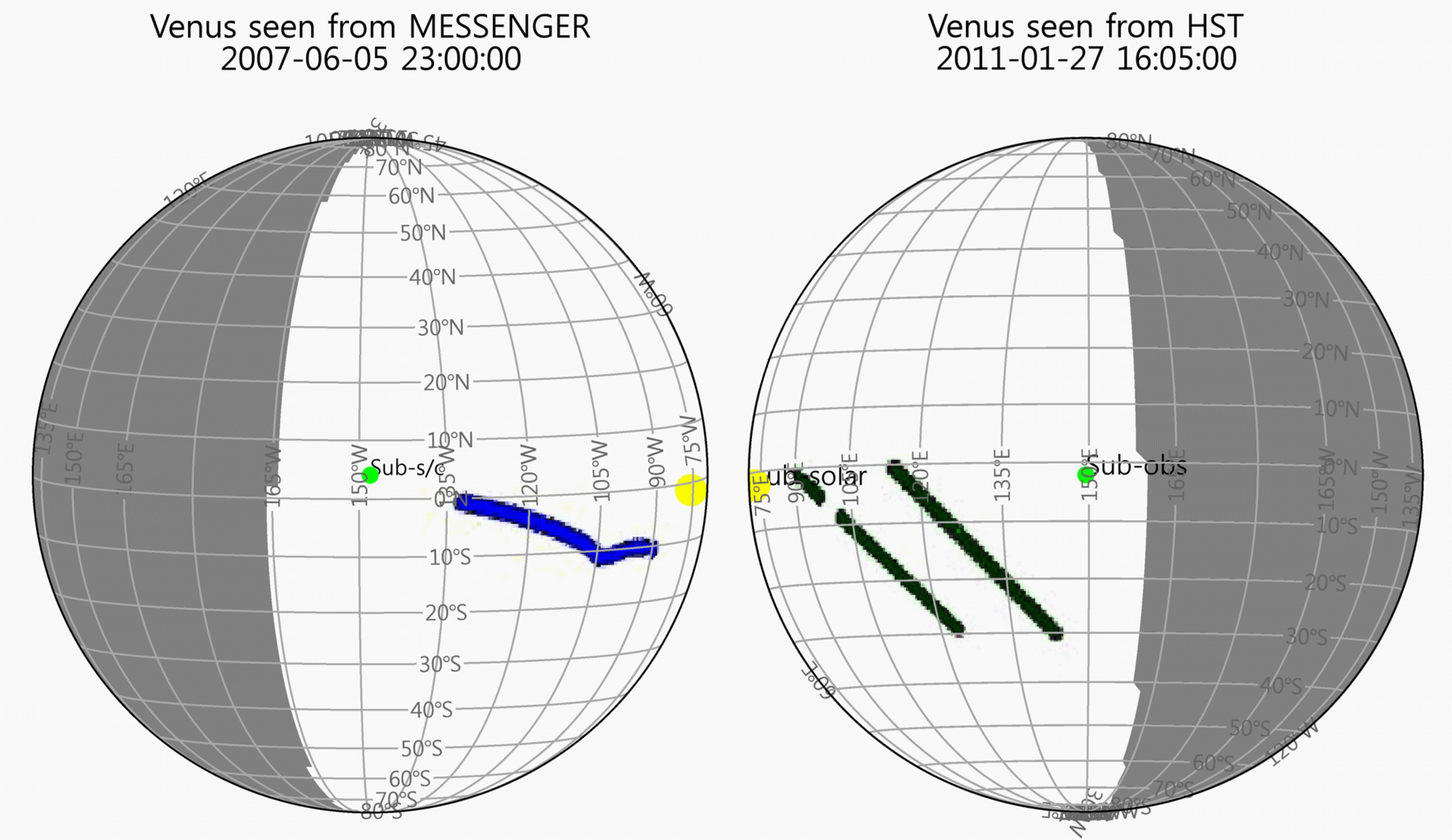}
\caption{Viewing geometries of spectral data: (left) MASCS on 2007 June 5 at the 72$^{\circ}$ of spacecraft-center of Venus-Sun (phase) angle position, while the marked blue data points were acquired at the 85--90$^{\circ}$ phase angles following the progress of the spacecraft during its flyby, and (right) STIS on 2011 January 27 as seen at 79$^{\circ}$ phase angle. White is dayside, and shaded area is nightside. Blue and dark green filled symbols indicate spectral data locations. Yellow symbols are a sub-solar point, and light green symbols are a sub-MESSENGER and a sub-Earth at the time indicated in the top of panels in UTC. \label{fig:globe_spec}}
\end{figure}

Except VMC, which mostly observed southern hemisphere \citep{Titov12}, the other three instruments observed both northern and southern hemispheres. For disk-resolved data, our comparison is done only for southern hemisphere, keeping the consistency with VMC data (see Section~\ref{subsubsec:disk-res}). Since 365-nm images show persistent dark low latitudes and bright high latitudes (figure~\ref{fig:img_comp}), we compare low (0--30$^{\circ}$) and high (50--70$^{\circ}$) latitudes separately for all images. Spectral data were taken using a narrow slit, and observed Venus across from local noon towards the terminator of a half illuminating phase, as shown in figure~\ref{fig:globe_spec}. All spectral data sets fall into low and middle (30--50$^{\circ}$) latitudinal bins, and we use only the southern low latitude bin data in this study. For disk-integration (Section~\ref{subsec:whole-diskA}), we did not distinguish hemispheres, and used all valid pixels on Venus disk. Table~\ref{tab:data} shows the configurations of four data sets used in this study.

\begin{table}
\renewcommand{\thetable}{\arabic{table}}
\centering
\caption{Summary of 365-nm observations used in this study. \label{tab:data}}
\begin{tabular}{c|c|cc|c|c}
\hline 
Instrument & VMC & \multicolumn{2}{c|}{UVI} & MASCS & STIS \\ 
Filter or channel & 365-nm & \multicolumn{2}{c|}{365-nm} & VIS-VIRS & G430L \\ 
\hline 
\multirow{2}{*}{Date} & 2006 May & \multicolumn{1}{c|}{2011 Feb.} & 2015 Dec. & 2007 & 2011 \\ 
 & $-$2014 May & \multicolumn{1}{c|}{$-$May} & $-$2017 May & Jun. 5 & Jan. 27 \\ \hline
Number of images/spectra & 17742 & \multicolumn{1}{c|}{82} & 1619 & 152 & 45 \\ 
Phase angle [$^{\circ}$] & 52.3$-$138.6 & \multicolumn{1}{c|}{0.9$-$62.9} & 0.9$-$144.0 & 89.7$-$90.0 & 79.3 \\ 
Latitude [$^{\circ}$] & Southern hemisphere & \multicolumn{1}{c|}{N \& S} &\multicolumn{1}{c|}{N \& S, or Southern} & 0$-$30S & 0$-$30S \\ 
Longitude [\edit2{$^{\circ}$E \deleted{E$^{\circ}$}}] & 0$-$360 & \multicolumn{1}{c|}{135-(360)-75} & 0$-$360 & 75$-$148 & 91$-$144 \\
\hline
\end{tabular} 
\end{table}

\section{Methods} \label{sec:methods}
For disk-resolved images, we \added{calculated a radiance factor \citep{Hapke12} (Section~\ref{subsubsec:reflectivity}), and then albedo, applying photometric correction (Section~\ref{subsec:photo-corr}).}\deleted{converted the observed radiance to a radiance factor, $r_{\rm F}$ \citep{Hapke12}, and then calculated albedo, applying photometric correction as described in \citet{Lee15a,Lee17}.} Spectral data were convolved using the filter transmittance function of the 365-nm channel of VMC, and \edit2{we applied the same photometric correction as for the images}. In order to take into account UVI's 2011 data, we calculated whole-disk albedo \citep{Sromovsky01,Garciamunoz14,Garciamunoz17}, without photometric correction \added{ due to the small apparent size of Venus (Section~\ref{subsec:whole-diskA})}. Radiative transfer model calculations were performed using the model and gaseous database in \citet{Lee15b,Lee16} \added{(Section~\ref{subsec:RTM}), to estimate the abundance of the unknown absorber that explains the observed 365-nm albedo, and to calculate solar heating rates}.

\subsection{Disk-resolved albedo} \label{subsec:disk-res_albedo}

\subsubsection{Reflectivity}\label{subsubsec:reflectivity}
\added{We converted the observed radiance to a radiance factor, $r_{\rm F}$, \citep{Hapke12}, the ratio of bidirectional reflectivity of a surface to the perfectly diffuse Lambertian surface illuminated normally. % citep{Lee15a}. 
We calculate the average solar flux at 1 AU \citep{Chance10} at each of UV filters of VMC and UVI, $S_{\odot}$ (W m$^{-2} \mu$m$^{-1}$), and the radiance factor as
\begin{equation}
\centering
(r_{\rm F})= \pi \beta (R_{\rm obs}) \frac {{d_{\rm V}}^2}{S_{\odot}},
\label{eq:r_F}
\end{equation}
where $R_{\rm obs}$ is the observed radiance (W m$^{-2}$ sr$^{-1} \mu$m$^{-1}$), $\beta$ is a calibration correction factor of VMC or UVI, and $d_{\rm V}$ is the distance of Venus to the Sun (in AU).}

%\subsection{Photometric corrections} \label{subsec:photo-corr}
\subsubsection{Photometric corrections} \label{subsec:photo-corr}
365-nm images show a combination of a smooth gradient from the sub-solar point to the terminator, and dark features owing to the presence of the unknown absorber. The smooth gradient depends on the incidence ($i$), emergence ($e$), and phase ($\alpha$) angles, which can be described by a photometric law (disk function), $D(\mu, \mu_0, \alpha)$, where $\mu=\cos(e)$ and $\mu_0=\cos(i)$. We can separate albedo, $A(\alpha)$, and disk function, $D(\mu, \mu_0, \alpha)$, from the radiance factor ($r_{\rm F}$) that is derived from the measured radiance \citep{Shkuratov11},
\begin{equation}
\centering
r_{\rm F}=A(\alpha)D(\mu, \mu_0, \alpha).
\label{eq:RF_AD}
\end{equation}
This albedo $A$, the equigonal albedo, depends on $\alpha$ \deleted{, and equals to the normal albedo when $\alpha=0$} \citep{Shkuratov11}.

Previous studies showed that the Lambert and Lommel-Seeliger law (LLS) performs better in describing the gradient depending on geometric angles, compared to the Lambert law and the Minnaert laws \citep{Lee15a,Lee17}. Therefore, in this study we adopted the LLS \citep{Buratti83,McEwen86}, 
\begin{equation}
\centering
D_{\rm LLS}=k(\alpha)\frac{2\mu_0}{\mu_0+\mu}+(1 - k(\alpha))\mu_0,
\label{eq:D_LLS}
\end{equation}
where $k$ is a coefficient depending on $\alpha$,
\begin{equation}
k(\alpha)=0.216004+0.00194196\times\alpha-(2.11589\times10^{-5})\times\alpha^{2},
\label{eq:k_LLS}
\end{equation}
and $\alpha$ is in [degree]. Eq.~\ref{eq:k_LLS} is updated from \citet{Lee17}, using more images to find the mean condition along phase angle.

%Disk-resolved data
\subsubsection{Areas of disk-resolved albedo for a comparison}\label{subsubsec:disk-res}
\begin{figure}
\plotone{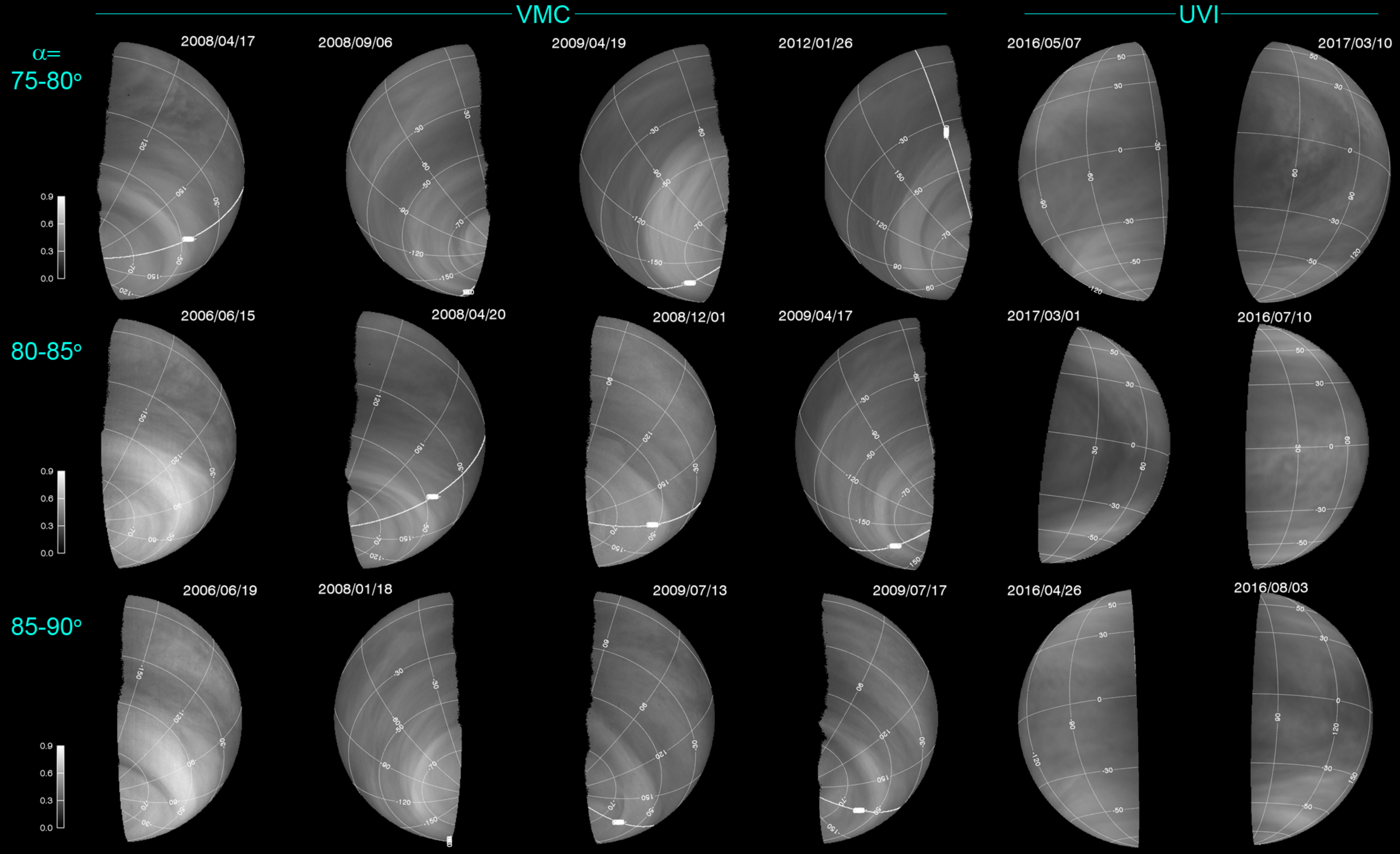}
\caption{Example 365-nm images of Venus. These present the cloud top albedo as observed by VMC and UVI between 2006 and 2017 when the observational phase angle was comparable to either STIS or MASCS Venus observations. In these example images, only Venus’ dayside a.m. or p.m. quadrant was visible within the camera field of view. First row shows images obtained in the 75--80$^{\circ}$ phase angle range, second row images were obtained at 80--85$^{\circ}$, and third images were obtained at 85--90$^{\circ}$. Left four columns are images taken by VMC covering latitudes extending from 90$^{\circ}$S to $\sim$10$^{\circ}$S, and the right two columns are pole-to-pole dayside images obtained by UVI. Though the higher latitudes are brighter in the UVI observations than the equator, the intense polar hood brightening detected at 40--70$^{\circ}$S by VMC in 2006 has not been observed by UVI. All images are photometrically corrected (see \replaced{Methods}{Section~\ref{subsec:photo-corr}}), and share the same color bar on the left.\label{fig:img_comp}}
\end{figure}
Figure~\ref{fig:img_comp} shows albedo $A$ at three 5-degree phase angle bins: 75--80, 80--85, and 85--90$^{\circ}$, from top to bottom. Left four columns are VMC images, having middle-latitudinal views, close to UVI's equatorial views on the right two columns. While UVI data have quality navigation using limb-fitting \citep{Ogohara17}, VMC data do not. So we restricted VMC data satisfying $i<84^{\circ}$, $e<81^{\circ}$, and $r_{\rm F}>0.05$ \citep{Lee15a}. The last condition causes non-smooth terminator for VMC images in figure~\ref{fig:img_comp}. As shown in these example images, we find no systematic tendency of albedo along local time, but temporal variations in the brightness and morphology. We also attempted to search for systematic variations in the albedo along longitude, as the surface topography may affect the 365-nm albedo, particularly over Aphrodite Terra \citep{Bertaux16}, but the longitudinal coverage of the VMC data is not evenly distributed over time, and additionally this depends on phase angle selections. A detailed analysis along longitude, latitude and time, requires a different approach from the broad range average utilized in this study. Here we focus on temporal variations using data obtained over a broad range of longitudes. We derive the mean latitudinal albedo from disk-resolved \added{VMC and UVI} images, \replaced{dividing}{divided into} two broad latitude bins: low (0-30$^{\circ}$) and high (50-70$^{\circ}$) latitudes. We derive a low latitudinal albedo also from the \added{MASCS and STIS} spectral data to complete the cross comparison \added{with VMC data} (Section~\ref{subsubsec:CrossComp_Disk-resA}).

\begin{figure}
\plotone{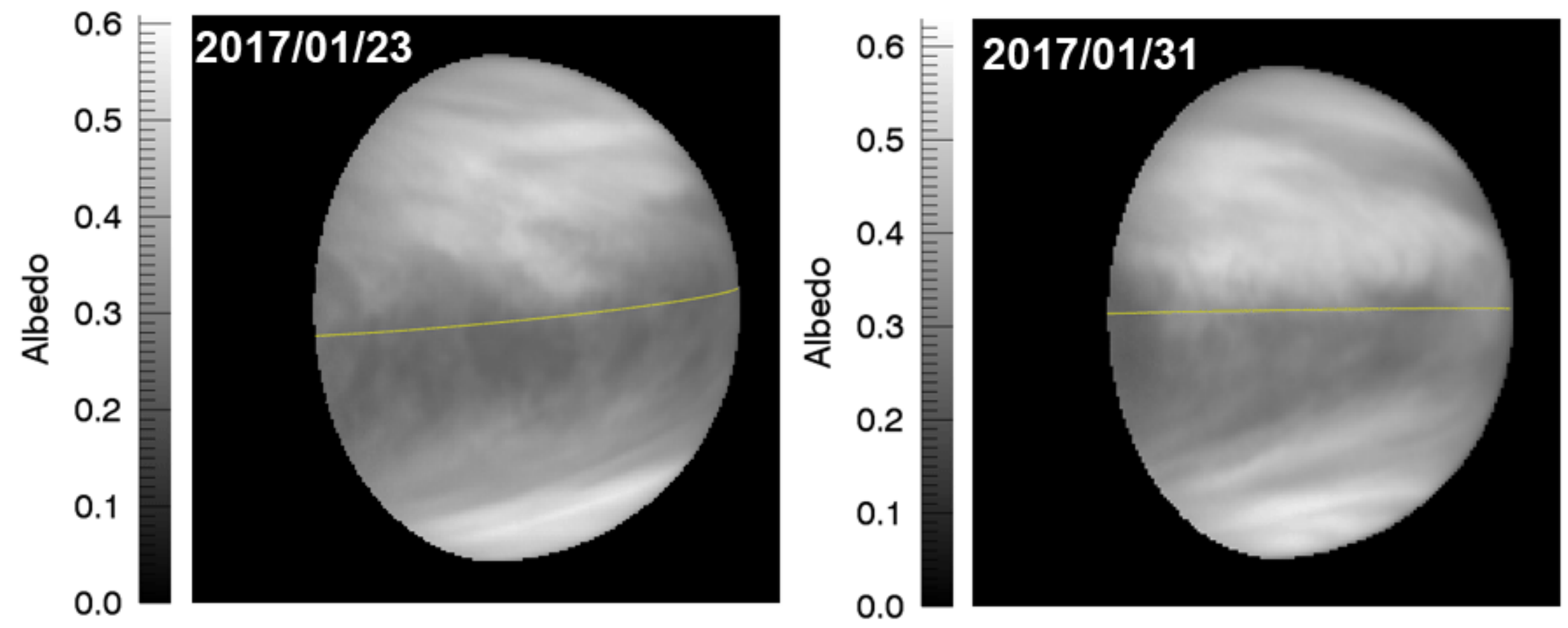}
\caption{Occasionally appearing hemispheric asymmetry of \added{365-nm} albedo across equator, observed on 2017 January 23 (left) and 31 (right) \added{using UVI on board Akatsuki}. Yellow line is the equator, and north is up. \label{fig:img_hasym}}
\end{figure}

From the equatorial orbit, UVI images show that cloud top albedo and contrast patterns primarily displayed north-south symmetry. However, as the example in figure~\ref{fig:img_hasym} shows, we also observed cloud top albedo patterns that were asymmetric across the equator. This asymmetry was observed frequently in January-February 2017. Wind fields retrieved from cloud motions also detected the similar asymmetry between northern and southern hemispheres over the same period \citep{Horinouchi18}. Thus, we restrict our disk-resolved data comparison only for southern hemisphere, which was observed by all of four instruments.

\subsubsection{Cross-comparison of disk-resolved VMC, MASCS, STIS, and UVI data} \label{subsubsec:CrossComp_Disk-resA}
The observed long-term decrease of albedo had been previously attributed to the sensitivity degradation \edit2{of VMC }by \citet{Shalygina15}. Following these authors, we used the 2.32 calibration correction factor (their Fig.12), which was retrieved from the comparison with VIRTIS-IR, and their value for the degradation ratio, $k_d$ ($=-16.2\times10^{-5}$ orbit$^{-1}$, and \edit2{1~orbit of Venus Express equals 1~Earth day\deleted{equals to the 1~Earth day}}). We correct all VMC data to the value at an initial sensitivity condition (0 orbit number of Venus Express) using the below equations given in \citet{Shalygina15}.
\begin{equation}
\centering
B(t)=B_0(t)\beta,
\label{B_t}
\end{equation}
where $t$ is time (orbit), $B$ and $B_0$ are the corrected and observed radiance respectively (W m$^{-2}$ sr$^{-1}$ $\mu$m$^{-1}$), and $\beta$ is the 2.32 calibration correction factor. Temporal sensitivity degradation correction is given as 
\begin{equation}
\centering
B(t_2)=B(t_1)\left(\frac{1+k_d t_2}{1+k_d t_1}\right),
\label{eq:B_corr}
\end{equation}
where $k_d$ is the sensor degradation factor (orbit$^{-1}$). We can get $B(t=0)$ as
\begin{equation}
\centering
B(t=0)=B(t)\left(\frac{1}{1+k_d t}\right).
\label{eq:B_0}
\end{equation}
So the final form is 
\begin{equation}
\centering
B(t=0)=B_0(t)\beta\left(\frac{1}{1+k_d t}\right).
\label{eq:B_0_end}
\end{equation}

\begin{figure}
\plottwo{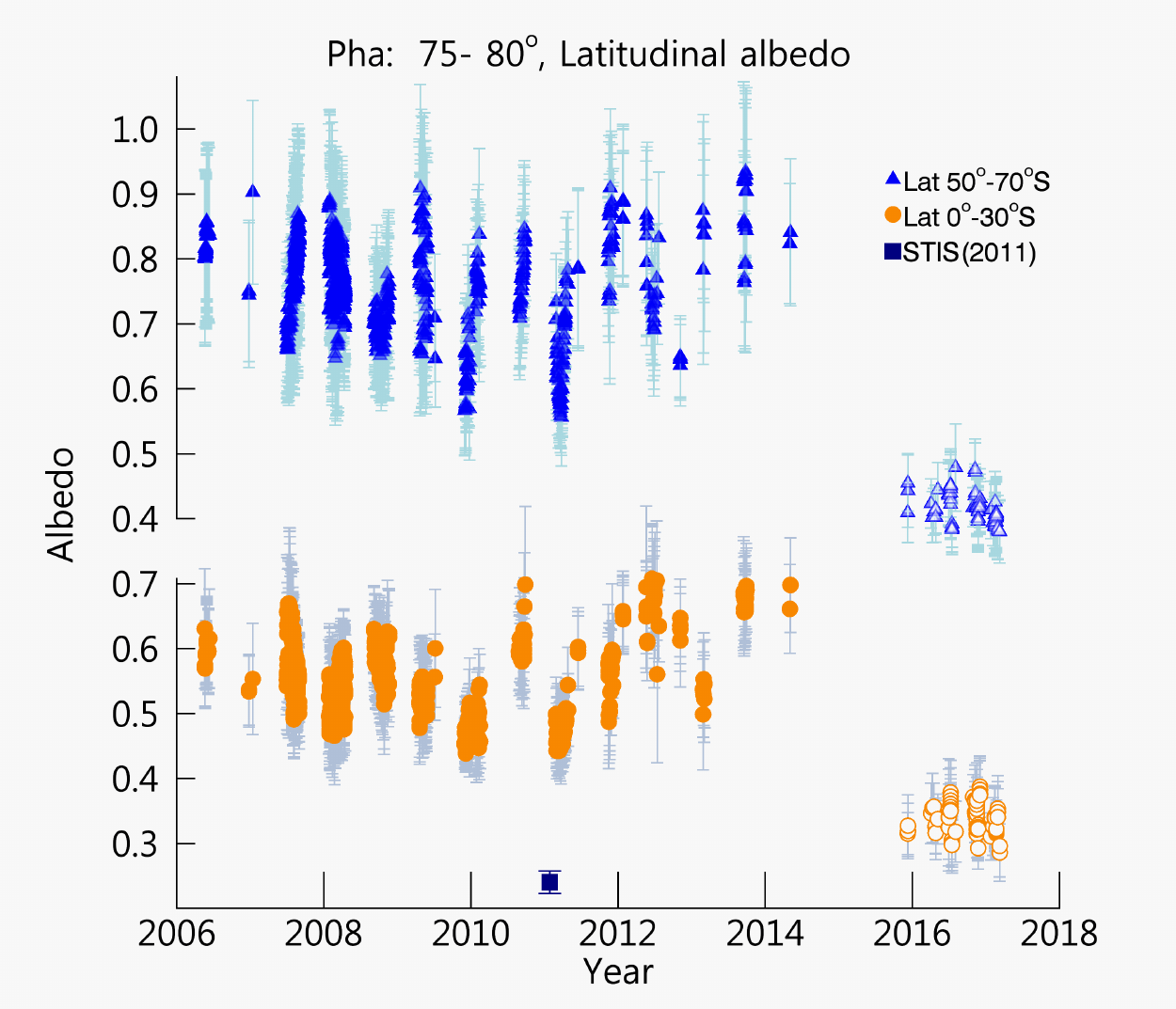}{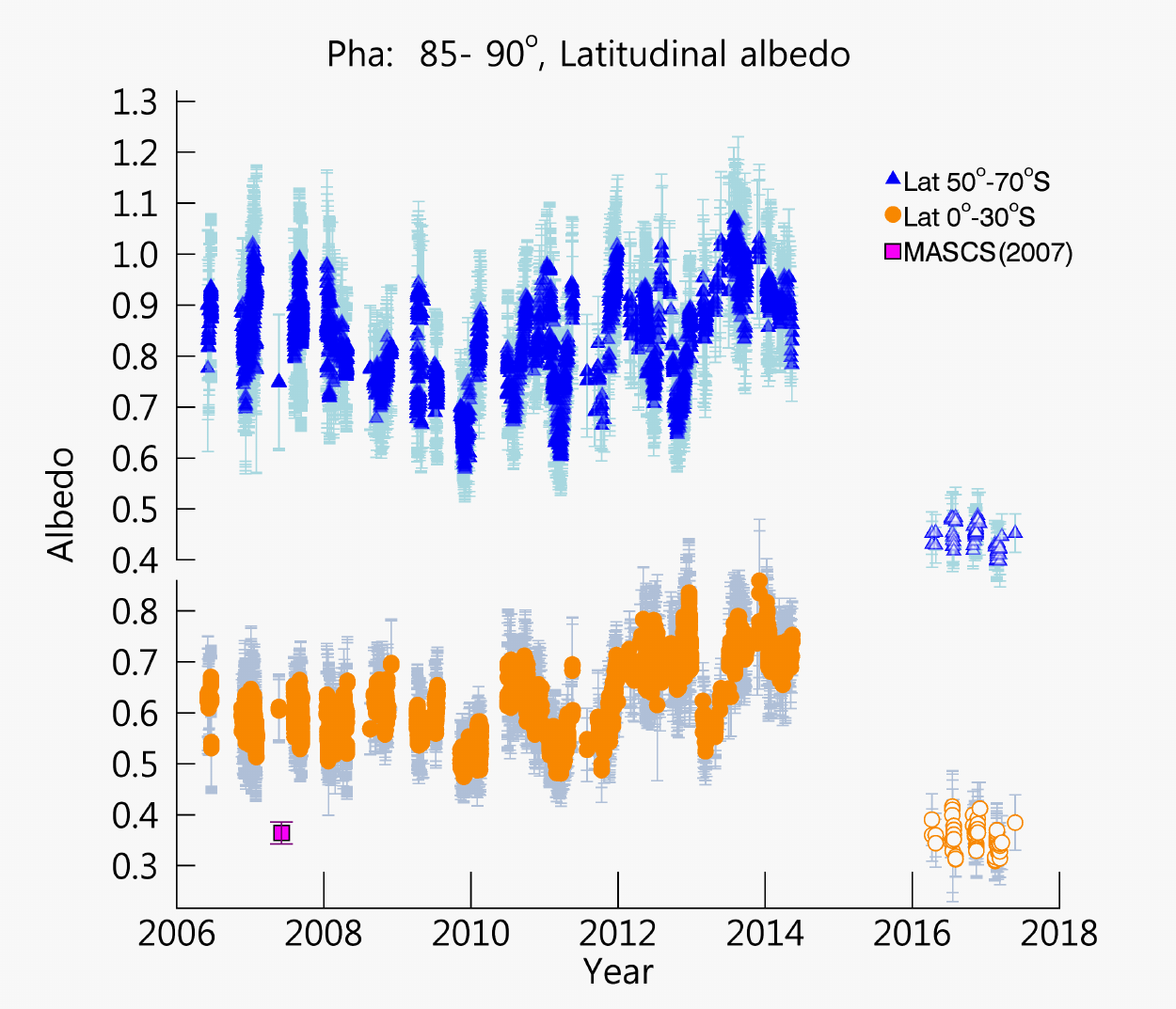}
\caption{Discarded results on 365 nm albedo comparison. The 365-nm albedo observed by VMC is corrected using Eq.~\ref{eq:B_0_end} \citep{Shalygina15} (filled symbols). Other 365-nm albedo observed by UVI (open symbols), STIS (navy square), and MASCS (magenta square) are compared. The low (0--30$^{\circ}$S) and high (50--70$^{\circ}$S) latitudinal mean albedo observed at 75--80$^{\circ}$ and 85--90$^{\circ}$ phase angle are shown in panels (Left) and (Right), respectively. Error bars are standard deviations of albedo. \label{fig:pha78N88_discarded}}
\end{figure}

Figure~\ref{fig:pha78N88_discarded} shows a comparison of low and high latitudinal albedo at the same phase angle bins, using the corrected VMC data with Eq.~\ref{eq:B_0_end} \citep{Shalygina15}. These corrected VMC data are significantly brighter than any of the independently calibrated MASCS, STIS, and UVI observations. The large difference between 2006 VMC and 2016 UVI is especially noticeable, while data in 2006 supposed to have least sensitivity degradation. We therefore discard this correction on VMC data due to the inconsistency with other calibrated MASCS, STIS, and UVI data.

\begin{figure}
\plottwo{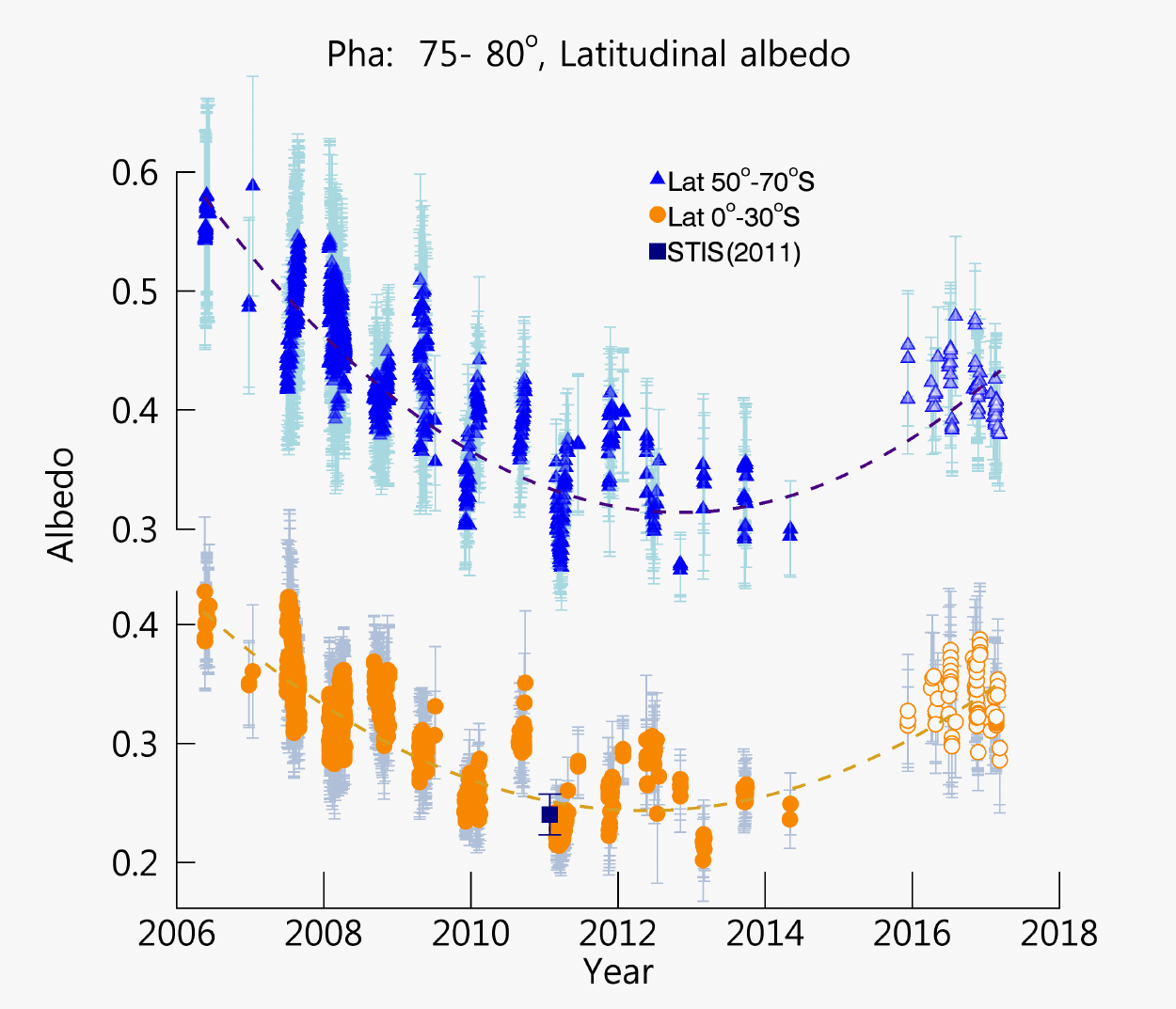}{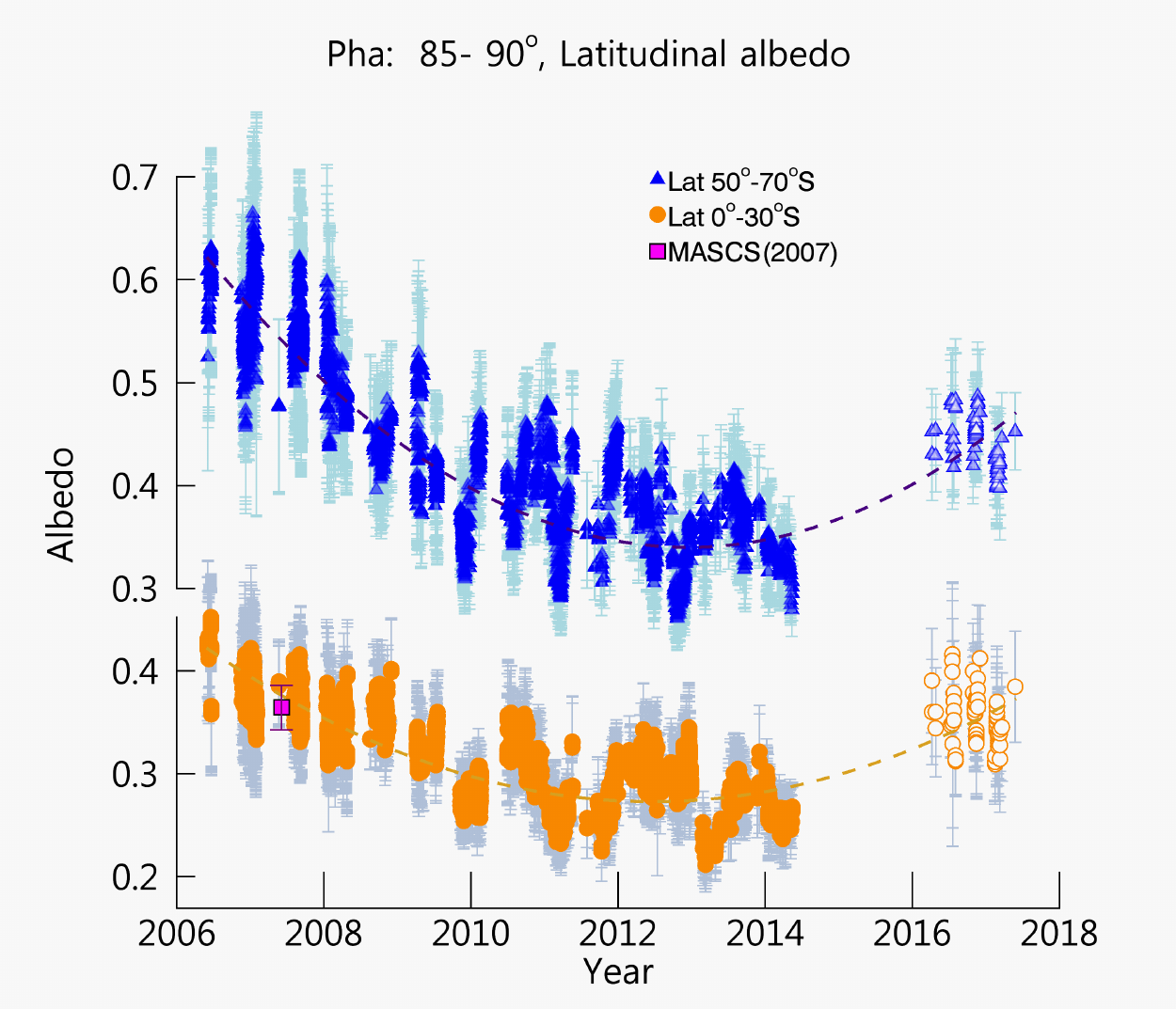}
\caption{Long-term variations of 365-nm albedo from 2006 to 2017 based on VMC (filled symbols), UVI (open symbols), STIS (navy square) and MASCS (magenta square). The low (0--30$^{\circ}$S) and high (50--70$^{\circ}$S) latitudinal mean albedo observed at 75--80$^{\circ}$ and 85--90$^{\circ}$ phase angle are shown in panels (Left) and (Right), respectively. A polynomial fit of temporal variation in the derived latitudinal mean albedo (long-dashed lines) highlights the overall temporal trend. Error bars are standard deviations of albedo.\label{fig:pha78N88}}
\end{figure}

Instead, we use the star calibration correction factor, 2.0$\pm$0.822, for the initial calibration correction factor ($\beta$) of the 365 nm channel of VMC \citep{Titov12,Shalygina15}. The large error of 82\% results in the ambiguous definition of the absolute radiance. In this study, we improve the calibration of the data using the data points of MASCS in 2007 and STIS in 2011 as a reference to fit VMC values (Table~\ref{tab:data}). To limit uncertainties that may arise from the influence of the aerosol scattering along phase angle ($\alpha$), we utilize only VMC data obtained at the same phase angle as either the MASCS ($\alpha=85-90^{\circ}$) or STIS ($\alpha=75-90^{\circ}$) observations. There are differences of 13 and 31 days from the closest VMC image at the time of the MASCS and STIS observations, respectively, at corresponding phase angles. To compensate for possible short-term fluctuations that may have occured during those time periods, we derive the 80-day mean albedo observed by VMC in the low-latitude bin at the two phase angle bins of MASCS and STIS. We then use the 80-day mean albedo to calculate ratios of $A$(MASCS)/$A$(VMC) at $\alpha=85-90^{\circ}$, and $A$(STIS)/$A$(VMC) at $\alpha=75-80^{\circ}$, where $A$ is mean low-latitude albedo. Using this process, the ratios of 0.74 for $A$(MASCS)/$A$(VMC) and 0.84 for $A$(STIS)/$A$(VMC) were inferred. Differences in the ratios may result from differences in the latitudinal coverage of the MASCS and STIS observations (figure~\ref{fig:globe_spec}), and may also incorporate possible temporal variation of sensitivity of VMC. Even though the latter is possible, the decreasing trend of low latitude albedo between 2007 and 2011 changes from 29\% to 24\% in the low latitudinal polynomial fit (figure~\ref{fig:pha78N88}), which is yet a minor effect on the results of this study. Using the mean value of the ratios, 0.79, over all the VMC data, the VMC and UVI albedo retrievals become reasonably comparable (figure~\ref{fig:pha78N88}). Thus, we adopt this value and apply a new modified VMC calibration correction factor of 1.58 ($\beta=2.0\times0.79$) to the VMC data used in our study. As figure~\ref{fig:pha78N88} shows, when the modified calibration correction factor is applied, the 365 nm albedo observed by VMC and UVI are reasonably aligned; those in 2008-2009 (VMC) are overlapped data in 2016 (UVI).

\subsection{Whole-disk albedo} \label{subsec:whole-diskA}
In order to evaluate robustness of the 1.58 modified VMC calibration correction factor, we employ 82 images taken with UVI in 2011 to compare with VMC. These UVI images were obtained after the first failure of the planned Venus orbit insertion of Akatsuki \citep{Nakamura14}. In those images, the apparent size of Venus is a few pixels across, but sufficient signal-to-noise ratios were achieved. We calculated the whole-disk albedo, $A_{\rm whole-disk}$, which is a function of phase angle ($\alpha$), following the equation below \citep{Sromovsky01},
\begin{equation}
\centering
A_{\rm whole-disk}(\alpha)=\frac{\pi}{\Omega_{\rm Venus}}\frac{{d_{\rm Venus}}^2F_{\rm Venus}}{S_{\odot}},
\label{eq:APhi}
\end{equation}
where $d_{\rm Venus}$ is the Venus distance to the Sun (AU), $F_{\rm Venus}$ is the measured disk-integrated Venus flux (W m$^{-2} \mu$m$^{-1}$), $\Omega_{\rm Venus}$ is the solid angle of Venus as seen from spacecraft (sr), and \replaced{$S_{\circ}$}{$S_{\odot}$} is the solar irradiance at 1 AU (W m$^{-2} \mu$m$^{-1}$)\edit2{, which is calculated for either UVI or VMC, using each of transmittance functions. \deleted{considering the each of filter transmittances of UVI and VMC}}.

The observed solid angle of Venus, $\Omega_{\rm Venus}$, is calculated as
\begin{equation}
\centering
\Omega_{\rm Venus}=\pi \left(\sin^{-1}\left(\frac{r_{\rm Venus}}{d_{\rm V-sc}}\right)\right)^2,
\label{eq:Omega_venus}
\end{equation}
where  $r_{\rm Venus}$ is the cloud top level radius of Venus (6052+70~km), and $d_{\rm V-sc}$ is the distance of spacecraft from Venus (km).
Observed Venus flux, $F_{\rm Venus}$, is calculated as
\begin{equation}
\centering
F_{\rm Venus}=\sum_{r<r_o}R_{\rm obs}(x,y)\times\Omega_{\rm pix},
\label{eq:F_venus}
\end{equation}
where $(x,y)$ is a location of \replaced{image in pixel}{pixel in image}, $\Omega_{\rm pix}$ is a solid angle of one pixel of either VMC or UVI, \replaced{$I_{\rm obs}$}{$R_{\rm obs}$} is radiance (W m$^{-2}$ sr$^{-1}$ $\mu$m$^{-1}$) in the target area ($r<r_o$), and $r$ is the distance of ($x,y$) from the Venus disk center (emission angle 0$^{\circ}$). $r_o$ is the radius range in which the measured radiance are summed considering the point spread function of the instrument (5~pixels) that is the required radius of aperture photometry of UVI star flux analysis. The whole-disk albedo can be expressed as $A_g\Phi(\alpha)$, where $A_g$ is geometric albedo, and $\Phi(\alpha)$ is the phase law of Venus\added{, describing the disk-integrated scattering efficiency as a function of phase angle} \citep{Garciamunoz14,Garciamunoz17}.

\subsubsection{Comparison of whole-disk albedo of VMC and UVI}\label{subsubsec:comp_whole-diskA}
We calculated the whole-disk albedo, $A_{\rm whole-disk}$ (Eq.~\ref{eq:APhi}), of the 82 UVI images in 2011, and all of the disk-resolved VMC and UVI images. The latter is possible because our analysis is restricted \replaced{with images having all day side along different solar phase, without missing any parts of day side due to}{to the images for which the observable dayside, defined by the observation elongation angle, is fully captured within} the field of view of cameras (see figure~\ref{fig:img_comp}). The 1.58 modified calibration correction factor is applied to the VMC data.

\begin{figure}
\plotone{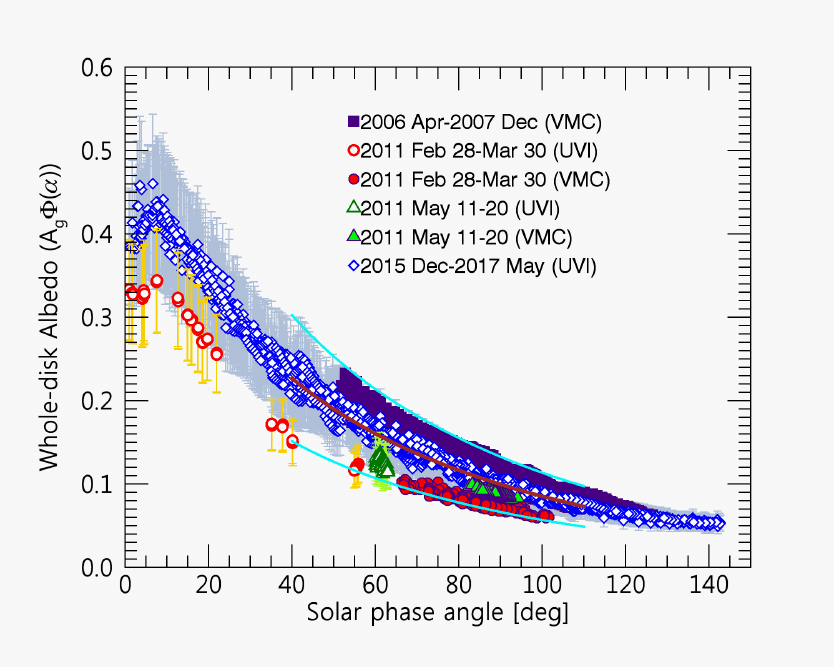}
\caption{Whole-disk albedo observed over a range of phase angles by VMC in 2006--2007 and 2011 and by UVI in 2011 and 2015--2017. The error bar of UVI is the measurement error. The solid brown line is the polynomial fit to the 2015 Dec -- 2017 May UVI data observed at phase angles of 40 to 110$^{\circ}$; the solid cyan lines highlight temporal albedo variations from 2006--2007 (upper line) to 2011 (lower line), where the relative shape of the phase angle dependence is assumed constant. The lower cyan line demonstrates successfully that UVI and VMC data in 2011 Feb--Mar are aligned on one phase curve.\label{fig:diskint_phase}}
\end{figure}

Figure~\ref{fig:diskint_phase} shows the results of the whole-disk albedo \replaced{along}{versus} phase angle. The 82 UVI images taken in February--March (circles) and in May 11--20 (triangles), are compared to the VMC images obtained contemporaneously. As a reference, the whole-disk albedo results derived from 2015--2017 UVI data and 2006--2007 VMC data are also included in the plot. The vertical bar of UVI data indicates the 18\% error in the absolute radiance (Section~\ref{sec:data}). Fractions of disk illuminated by the Sun changes \replaced{along}{with} the solar phase angle, from 100\% at 0$^{\circ}$ phase angle to 0\% at 180$^{\circ}$ phase angle, so there is a dominant decreasing trend as phase angle increases. At small phase angles, a local minimum related to glory is apparent \citep{Garciamunoz14,Lee17}. The polynomial fit of UVI's 2015--2017 data shows the empirical phase function $\Phi(\alpha)$ in the 40--110$^{\circ}$ phase angle range (brown solid line, Eq. is shown in Table~\ref{tab:poly_ci}). We shift this phase function vertically to fit the maximum and minimum VMC whole-disk albedo at 75--80$^{\circ}$ phase angle (cyan lines), assuming a change of geometric albedo $A_g$ over time whereas the scattering properties $\Phi$ are the same. The lower cyan line that fits the VMC whole-disk albedo in 2011 February--March, encompasses UVI data at the same period. This means that the 1.58 modified calibration correction factor for VMC data works well.

In addition, our analysis of the 2011 and 2015--2017 UVI whole-disk albedo at 0$^{\circ}$ phase angle, which is a geometric albedo $A_g$, increases from $\sim$0.33 in 2011 to $\sim$0.40 in 2015--2017. This result confirms the increasing albedo trends \edit2{from 2011 to 2015--2017, }observed at high and low latitudes based on the \edit2{recalibrated disk-resolved 2011 VMC and disk-resolved 2015--2017 UVI data\deleted{disk-resolved UVI (2015-2017) and recalibrated disk-resolved 2011 VMC data}} (figure~\ref{fig:pha78N88}). The observation of an increasing albedo over these time periods completely opposes the behavior that would be produced by progressive long-term UV sensor degradation \citep{Shalygina15}. This cannot be used to investigate the influence of surface topography on the 365-nm albedo because for each date a broad range of longitudes is included in the \replaced{deviation}{derivation} of the albedo at small phase angles\added{; this includes the }\deleted{(}150--315\edit2{$^{\circ}$E} range in 2011 March, \added{the} 180--330\edit2{$^{\circ}$E} in 2016 May, \added{and the} 165--(360)--45\edit2{$^{\circ}$E} \added{range} in 2016 December--2017 January\deleted{)}. Additionally, the 2011 UV data have insufficient spatial resolution to segregate \deleted{a }specific longitude and latitude topographic regions.

\subsection{Radiative transfer model} \label{subsec:RTM}
\begin{figure}
\plotone{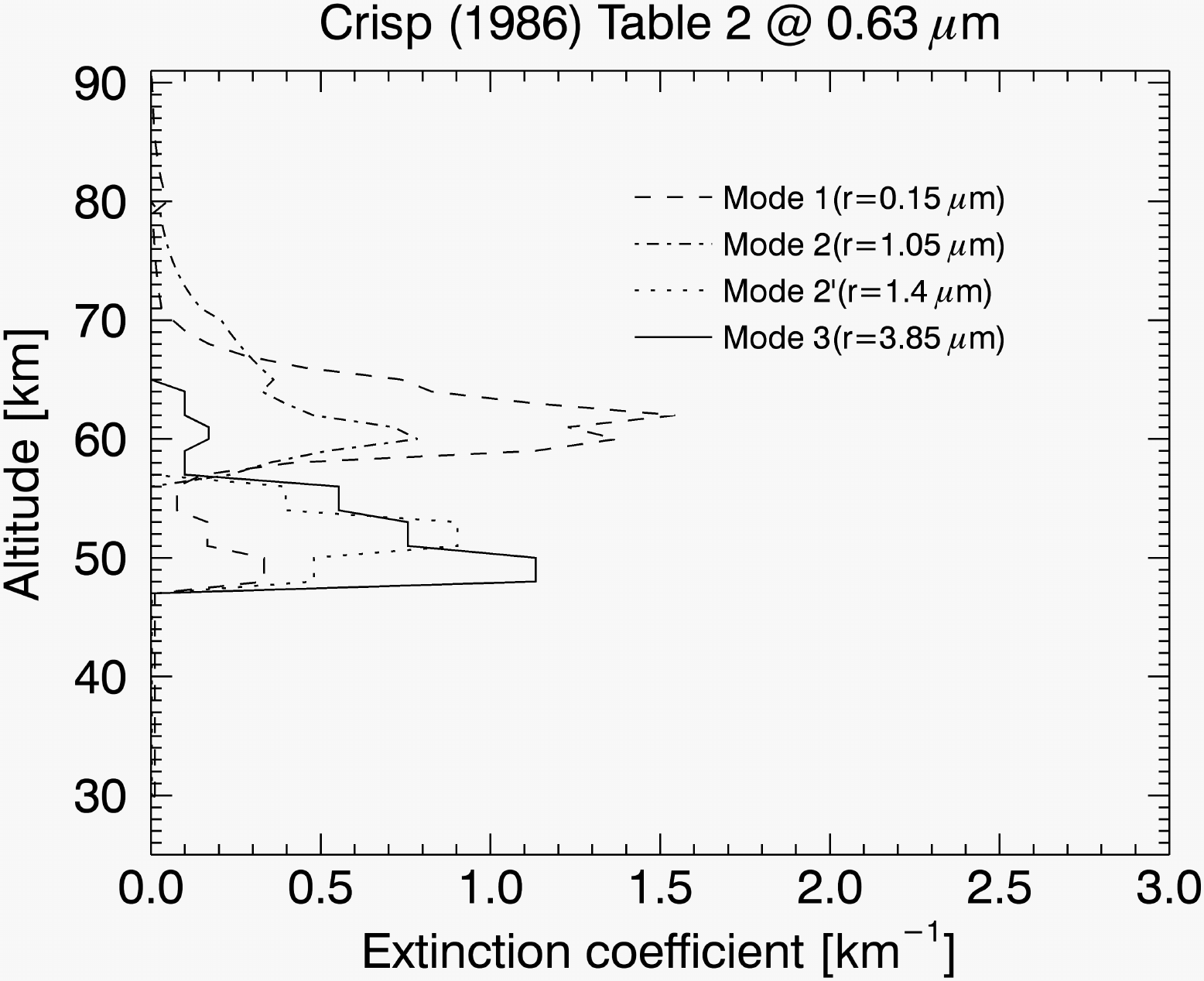}
\caption{Vertical profile of the cloud extinction coefficient, used in this study.\label{fig:cloud_ext}}
\end{figure}

We use a one-dimensional line-by-line radiative transfer model (SHDOM, \citeauthor{Evans98} \citeyear{Evans98}) in the 0--100~km altitude range and in the 2000--50000~cm$^{-1}$ ($=0.2-5~\mu$m) range to estimate the abundance of the unknown absorber that fits the observed 365-nm albedo, and to calculate the solar heating rate at low latitudes at the local noon time. The configurations are the same as used in \citet{Lee15b,Lee16}. CO$_2$ line parameters were taken from a combined HITEMP2010 \citep{Rothman10}, and one developed by \citet{Wattson92} and \citet{Pollack93}, as described in \citet{Lee16}. We included collision-induced CO$_2$ absorption in near-infrared, and that of H$_2$O \citep{Lee16}. Line parameters of other gases, N$_2$, SO$_2$, OCS, HCl, CO, HF, and H$_2$S, were taken from HITRAN2012 \citep{Rothman13}, and vertical profiles of gaseous abundances were taken from \citet{Titov07}. We included Rayleigh scattering \citep{Pollack67,Hansen74}, and UV range absorption cross sections of SO$_2$ \citep{Wu2000}. Microphysical properties of the cloud aerosols (mode 1, 2, 2’, and 3) were taken from \replaced{\citep{Zasova07}}{\citet{Zasova07}}. We took the vertical structures of clouds\added{'} extinction coefficient from \citet{Crisp86}\added{, as shown in figure~\ref{fig:cloud_ext}}. For the unknown absorber, we assumed \citet{Crisp86}'s absorption coefficient ($Q_{\rm abs}$) of the mode 1 particle (0.15-$\mu$m mean radius cloud particles)\added{ \citep{Knollenberg80,Kawabata80,Wilquet09,Luginin16}} in the 0.3--0.8~$\mu$m spectral range in the upper cloud layer (57--71~km). This assumed vertical location of the unknown absorber has been widely adopted in previous solar heating calculations \citep{Crisp86,Lee15b,Haus16}.

\section{Results} \label{sec:results}

\subsection{Temporal variations of low and high latitudinal albedo} \label{subsec:disk-resolved-results}
The albedo $A$ (Eq.~\ref{eq:APhi}) is phase angle $\alpha$ dependent due to strong backward and forward aerosol scatterings \citep{Lee15a,Shalygina15,Garciamunoz14}. Therefore, we restrict a comparison of $A(\alpha)$ to data obtained at near equivalent phase angle bins. Figure~\ref{fig:pha78N88} shows the temporal evolution of Venus' low and high latitudinal mean albedo obtained at phase angles of 75--80$^{\circ}$ and 85--90$^{\circ}$. In this figure a strong and steady decline in the albedo occurs between 2006 and 2011, from 0.4 to 0.25 at low latitudes and from 0.6 to 0.3 at high latitudes. Albedos at low and high latitudes in 2015 December--2017 May are restored to the 2008-2009 values.

\begin{figure}
\plottwo{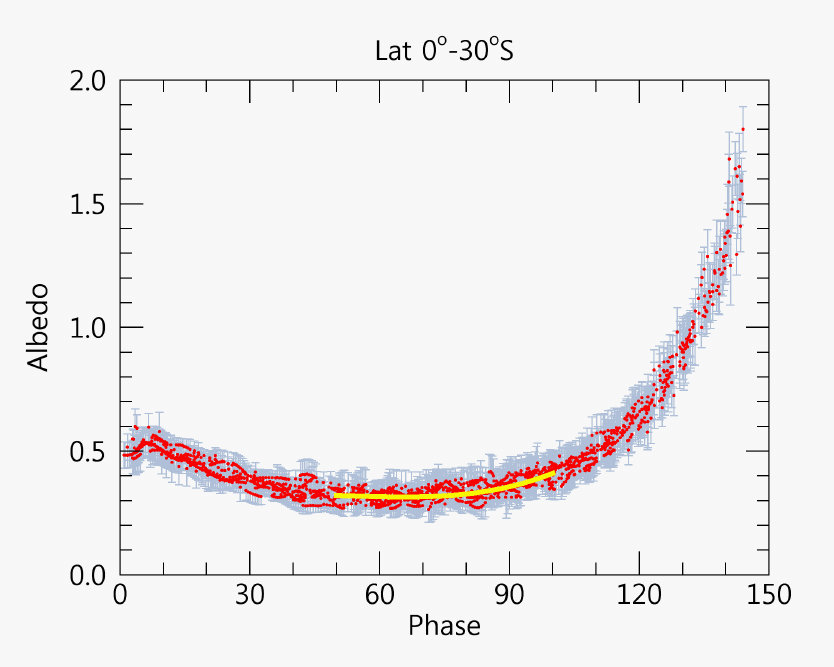}{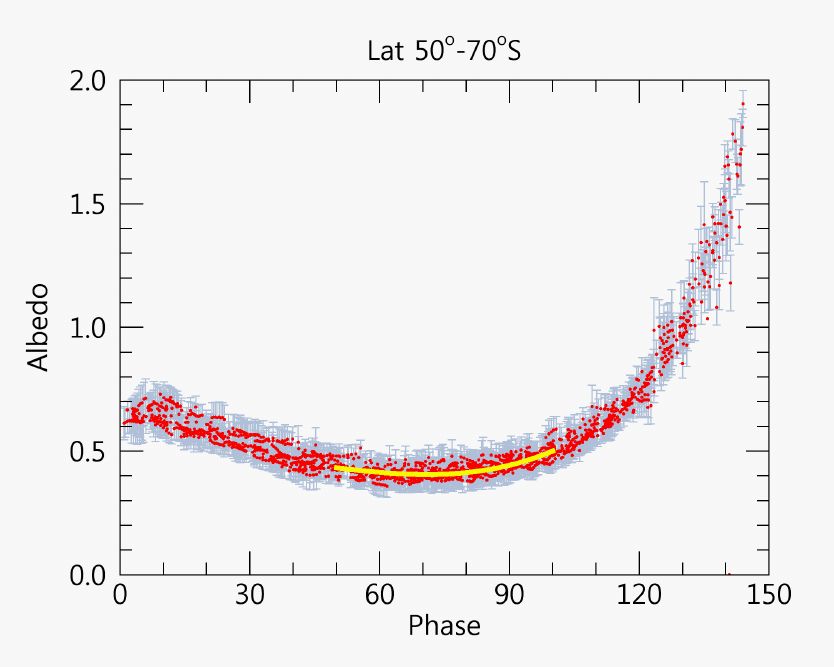}
\caption{Phase curve of mean $A$ at low (left) and high (right) latitudes in 2016--2017. Yellow lines are empirical polynomial fits. Vertical bars are standard deviations. \label{fig:EQ_CC_phasefn}} 
\end{figure}

While direct comparison of the mean $A$ should be done using data at a specific phase angle bin, often there are missing data over time due to regular changes in phase angles along the orbit of the spacecraft. 
In order to have a better temporal coverage of the mean $A$ variations, we calculate the percent deviation of $A$ from the 2016--2017 mean phase curve, $\bar{A}_{\rm UVI}(\alpha)$, which is derived as a polynomial fit of UVI data in 2016--2017 in the 50--100$^{\circ}$ phase angle range (figure~\ref{fig:EQ_CC_phasefn}). Table~\ref{tab:poly_ci} shows the coefficients of the polynomial fit to the mean phase curves at low and high latitudes. Deviations from the UVI phase curve are defined as:
\begin{equation}
\centering
{\rm Deviation~[\%]}=\frac{A(\alpha)-\bar{A}_{\rm UVI}(\alpha)}{\bar{A}_{\rm UVI}(\alpha)}\times100.
\label{eq:dev_A}
\end{equation}
\begin{table}
\renewcommand{\thetable}{\arabic{table}}
\centering
\caption{Coefficients of a polynomial fit, \edit2{$y(\alpha)=\sum_{i=0}^{n}{c_i}{\alpha}^i$, where $n\leq8$,}\deleted{$y(\alpha)=\sum_{i=0}^{i}{c_i}{\alpha}^i$} in Figures~\ref{fig:diskint_phase} and \ref{fig:EQ_CC_phasefn}. \label{tab:poly_ci}}
\begin{tabular}{c|ccc}
 & $\bar{A}_{\rm UVI, low-lat}(\alpha)$ & $\bar{A}_{\rm UVI, high-lat}(\alpha)$ & $\bar{A}_{\rm UVI, whole-disk}(\alpha)$ \\ 
\hline 
$c_0$ & 0.18708465 & \replaced{0.18708465}{0.47673713} & 0.62539023 \\
$c_1$ & 0.0086404233 & \replaced{0.0086404233}{0.0029037657} & $-$0.022833883 \\
$c_2$ & $-$0.00017491717 & \replaced{$-$0.00017491717}{$-$0.00012465654} & 0.00063482472 \\
$c_3$ & 1.1079428$\times10^{-6}$ & \replaced{1.1079428$\times10^{-6}$}{9.7921542$\times10^{-7}$} & $-$1.2777032$\times10^{-5}$ \\
$c_4$ &  &  & 1.7206082$\times10^{-7}$ \\
$c_5$ &  &  &  $-$1.4916523$\times10^{-9}$ \\
$c_6$ &  &  & 7.9181817$\times10^{-12}$ \\
$c_7$ &  &  & $-$2.3230007$\times10^{-14}$ \\
$c_8$ &  &  & 2.8698482$\times10^{-17}$ \\ 
\end{tabular} 
\end{table}
\begin{figure}
\plottwo{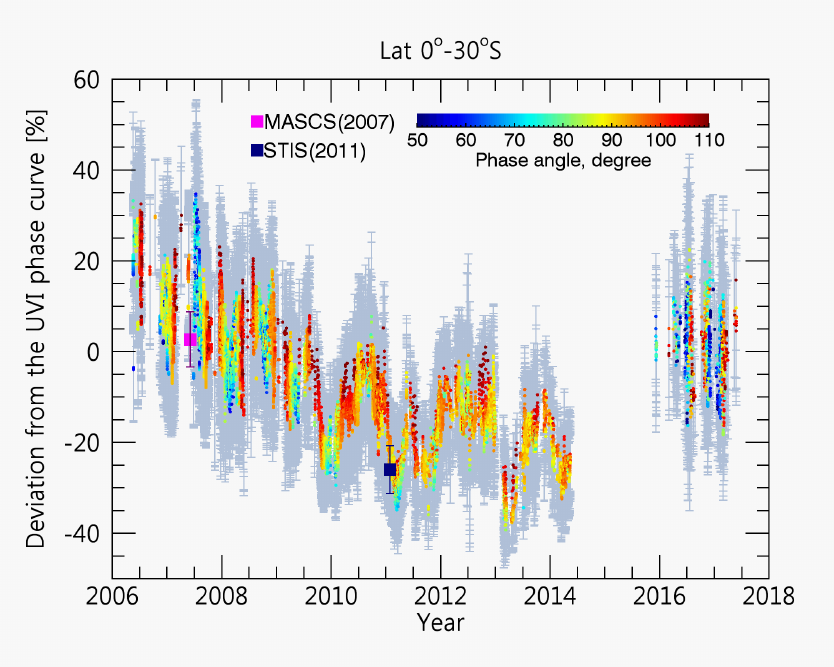}{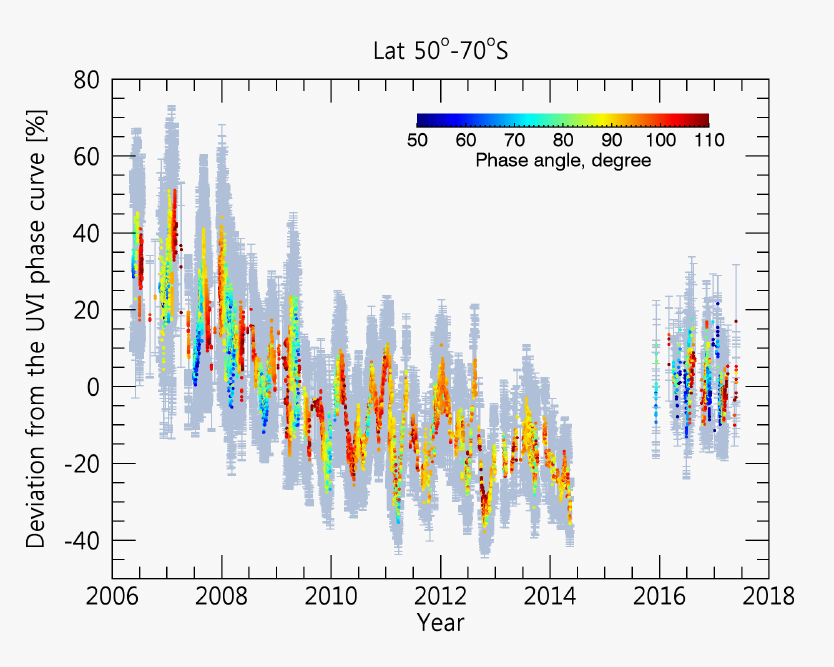}
\caption{Relative temporal variations of mean $A$ compared to the 2016--2017 phase curve, defined as Eq.~\ref{eq:dev_A}, at low (left) and high (right) latitudes. Vertical bars are standard deviations of mean $A$. The color of dot-symbols indicates phase angles between 50 and 110$^{\circ}$, shown in the color bar. MASCS (2007) and STIS (2011) at low latitudes are shown with square symbols. \label{fig:EQ_CC_devA}} 
\end{figure}

Figure~\ref{fig:EQ_CC_devA} shows percent of deviations of mean $A$ from $\bar{A}_{\rm UVI}(\alpha)$ at low and high latitudes as a function of time, where phase angles from 50 to 110$^{\circ}$ are represented according to the color bar at the top of each panel. The overall decline in the 365-nm albedo from 2006 to 2011 remains apparent. Additionally, relatively sharp albedo declines are observed at the end of 2009 and 2010, and at the beginning of 2013 that remain constant over short time periods ($\sim$months) before returning to the $\bar{A}_{\rm UVI}(\alpha)$ level. The robustness of the darker albedo conditions observed at the beginning of 2011 are confirmed by the overlap with the January 2011 STIS data. At high latitudes, periods of albedo decrease are less pronounced and appear to be shorter lived than those at low latitudes. This may be an indication of the combined influence of the unknown absorber abundance and the meridional circulation (Hadley circulation). In particular, the latter would remove older aerosols downward below the cloud top level, following the descending branch at high latitudes \citep{Imamura01}, and leaving behind only the bright newly formed aerosols that support the existence of Venus' bright high latitudes.

\subsection{Temporal variations of whole-disk albedo} \label{subsec:whole-disk-results}
\begin{figure}
\plotone{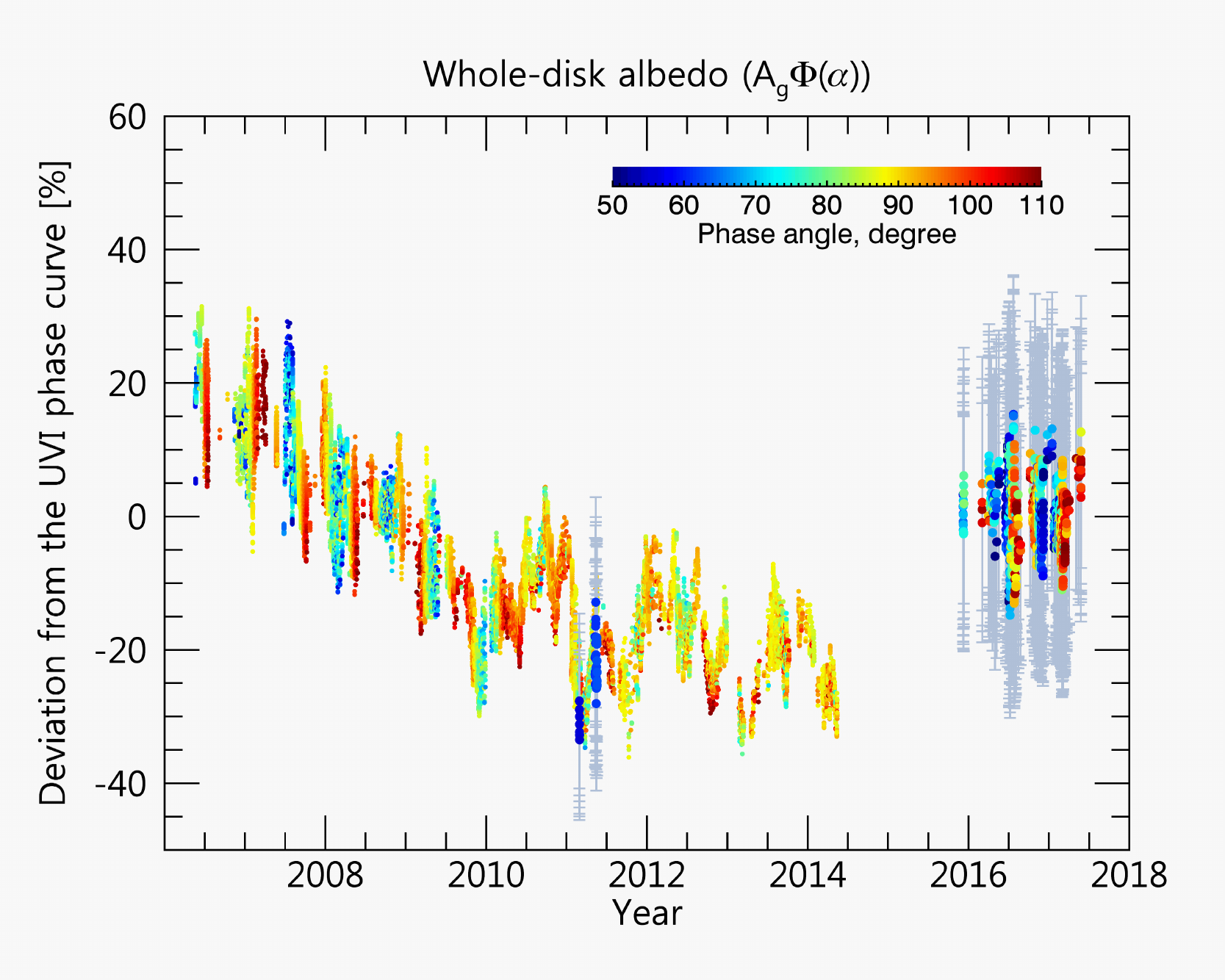}
\caption{Deviations of the whole-disk albedo, $A_{\rm whole-disk}$ of all VMC (2006--2014) and UVI (2011, 2015--2017) data. Vertical bars of UVI correspond to the 18\% error in the absolute radiance inferred from the star calibration. The symbol colors indicate phase angle, as defined in the color bar. \label{fig:wholeDiskA_dev}} 
\end{figure}

We apply Eq.~\ref{eq:dev_A} also for the whole-disk albedo $A_{\rm whole-disk}$ and use the mean phase curve (figure~\ref{fig:diskint_phase} and Table~\ref{tab:poly_ci}) to get the percent deviations of $A_{\rm whole-disk}$ from the 2016--2017 mean phase curve. Figure~\ref{fig:wholeDiskA_dev} shows the result as a function of time. The same albedo decreases inferred from the disk-resolved data from 2006 to 2011 are shown in this whole-disk albedo analysis. The independent UVI data obtained in early 2011 (dots with error bar) are well overlapped with those of VMC, including the same short-term albedo variations between March and May 2011. A darker 365-nm albedo in 2011 than 2006\edit2{--}2007 or 2016--2017 is a common feature in both the disk-resolved latitudinal and disk-integrated data, implying that the 365-nm darkening that occurred between those dates was a global phenomenon.

\subsection{Solar heating variations} \label{subsec:RTMresults}
Because of the influence of the unknown absorber on solar heating rate near the cloud top level \citep{Crisp86,Titov07,Lee15b} owing to the broad absorption spectrum from UV to visible \citep{Crisp86,Perezhoyos18}, the long-term 365 nm albedo variation we present here should have had a significant effect on Venus' solar heating rate. We use the radiative transfer model (Section~\ref{subsec:RTM}) to determine a required abundance of the unknown absorber, which is assumed to be fixed to the mode~1 particles in the 57--71~km altitude range (Section~\ref{subsec:RTM}), and to estimate the resulting solar heating rate changes at low latitudes. We incorporate the observed low latitudinal 365-nm albedo variations into our model using a scaling factor $f$ that is multiplied \replaced{to}{by} the assumed initial absorption coefficient $Q_{\rm abs}$ taken from \citet{Crisp86} for the unknown absorber. $f=1.0$ means the initial value. $f>1.0$ means more abundant unknown absorber, which results in darker albedo, and vice versa. So we control only the single scattering albedo of the mode 1 particles, but not the size of the particles. In order to fit the observed $A$, we calculate 5 sets of emission ($e$) and incident ($i$) angles that \replaced{satisfying}{satisfy} $\alpha=88^{\circ}$: $(e, i)=(28^{\circ}, 60^{\circ}), (38^{\circ}, 50^{\circ}), (48^{\circ}, 40^{\circ}), (58^{\circ}, 30^{\circ}),$ and $(68^{\circ}, 20^{\circ})$, to simulate different combinations of $e$ and $i$. The resulting radiance factors are corrected using the same photometric law which was applied to the calculation of the albedo (Section~\ref{subsec:photo-corr}), and the mean value is used to find $f$ values that match the observed albedo.

\begin{figure}
\plotone{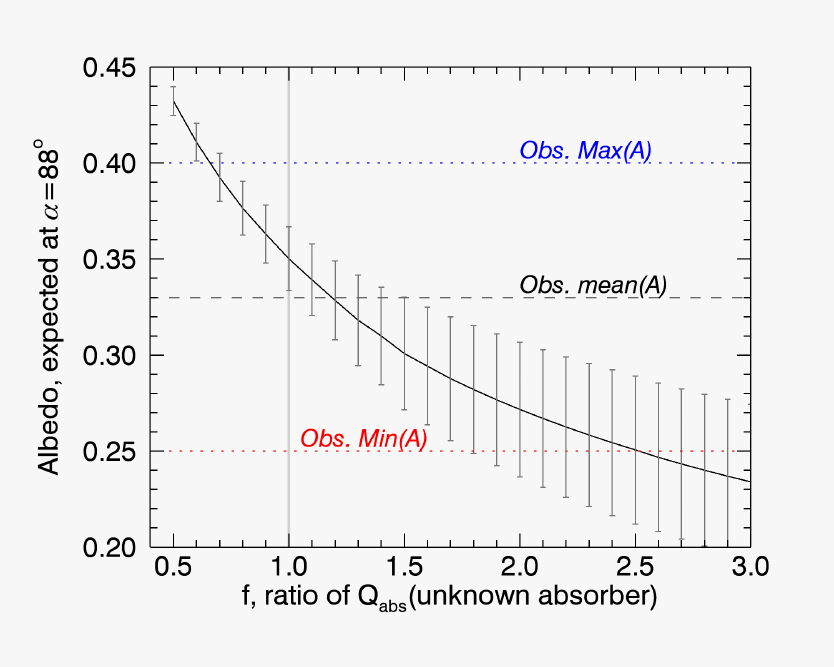}
\caption{Expected albedo as a function of $f$. Vertical error bars correspond to uncertainties depending on incidence and emission angles (see text for details). \label{fig:f}} 
\end{figure}

\begin{figure}
\includegraphics[height=0.7\paperheight]{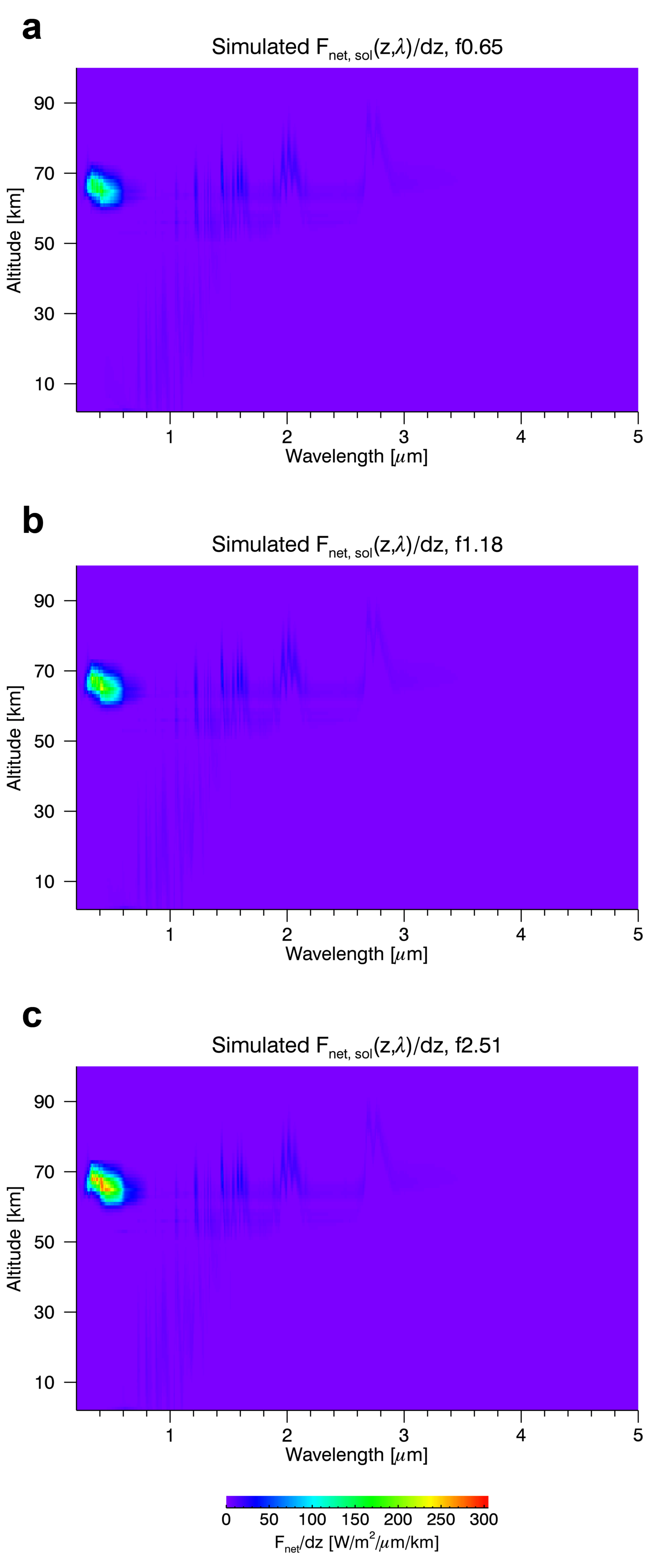}
\caption{Net solar flux divergence as functions of wavelength and altitude at local noon time at 15$^{\circ}$S. (a) f=0.65, (b) f=1.18, and (c) f=2.51 are multiplied \replaced{for}{by} the extinction coefficients of the unknown absorber ($Q_{\rm abs}$) in the 57--71~km altitude range and in the 0.3--0.8~$\mu$m range (see text for details). Spectral features are smoothed over 0.01~$\mu$m interval. \label{fig:divfnet}} 
\end{figure}

Figure~\ref{fig:f} shows the relationship of $f$ and the mean value of calculated 365-nm albedo for the 88$^{\circ}$ phase angle conditions (Figure~\ref{fig:pha78N88}). The mean albedo observed at low latitudes is 0.33, requires $f=1.18$. This is close to the initial value, $f=1.0$, that results in a calculated albedo $(A)=0.35$. The approximate maximum albedo observed at low latitudes, $A=0.40$, requires $f=0.65$, and the minimum albedo, $A=0.25$, requires $f=2.51$. We employ these $f$ values in net solar flux profile calculations at 15$^{\circ}$ latitude, which is the middle of the low latitude bin, at local noon time. Figure~\ref{fig:divfnet} shows the net solar flux divergence spectrum as a function of altitude $z$, $-\nabla \cdot F_{\rm net}=-({dF_{\rm net}}/{dz})$, 
%$-\frac{dF_{\rm net}}{dz}$
where $F_{\rm net}$ is net solar flux. The strongest influence of the unknown absorber appears around 400~nm, \deleted{at} where decreasing absorption and increasing solar irradiance \replaced{are overlapped}{overlap}.

\begin{figure}
\plottwo{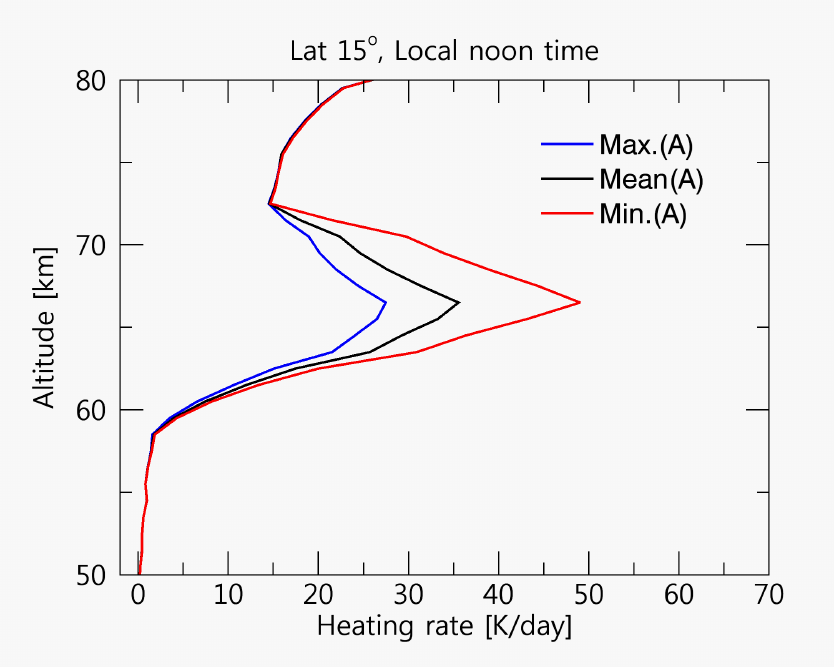}{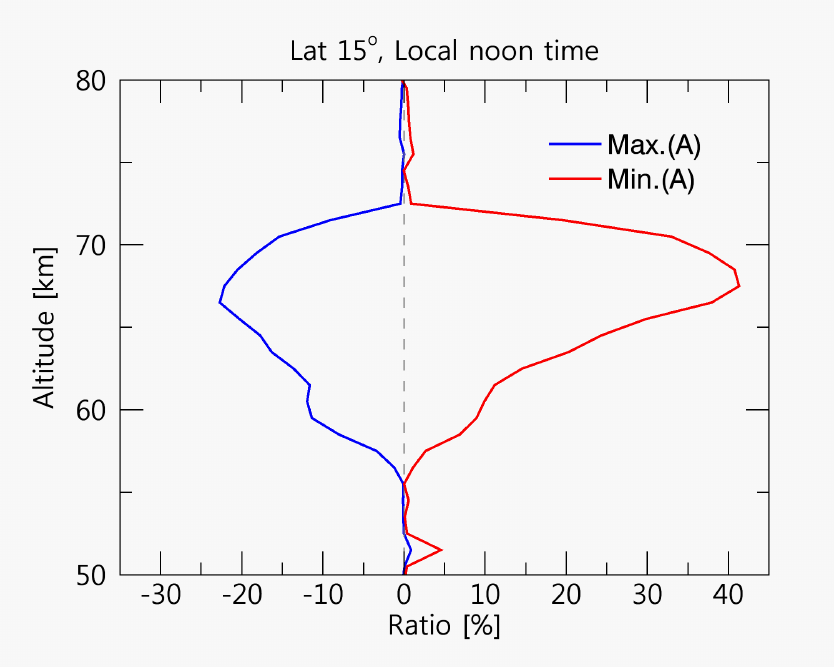}
\caption{Calculated solar heating rate profiles for the observed maximum albedo, minimum albedo, and mean albedo at local noon time at 15$^{\circ}$S. Solar heating rate is displayed in K/day (left), and as a relative ratio from the mean albedo (right). \label{fig:HR}} 
\end{figure}

Figure~\ref{fig:HR} shows the calculated local noon time solar heating rate at 15$^{\circ}$ latitude, derived from Figure~\ref{fig:divfnet}. This represents that the solar heating rate varied from \edit2{$-$}25\% to +40\% from the mean (2006--2017), which is a significantly large range. The peak of solar heating rate is 36~K/(Earth)day for the mean albedo condition, 49~K/day for the minimum albedo, and 27~K/day for the maximum albedo at this local noon time. 

We note that the vertical structure of the upper clouds may affect the solar heating rate \citep{Lee15b}, but the structure of the upper clouds at low latitudes from near infrared observations is shown to be rather stable during the time of Venus Express \citep{Ignatiev09,Cottini15,Fedorova16}, validating the use of a fixed cloud structure. Other analyses suggest that the vertical distribution of the unknown absorber may sometimes extend vertically above the cloud top level \citep{Molaverdikhani12,Lee15a}. However, the upper haze and vertical locations of the unknown absorber are not changed in our calculations as these are beyond the scope of this study. Further sensitivity studies on solar heating rate would explore in depth the effects of variable vertical and latitudinal distributions of the unknown absorber on Venus' global solar heating rate.

\section{Discussion} \label{sec:discussion}
\subsection{Relationship between solar heating and zonal winds at the cloud top level} \label{sec:dis1_solarheating-winds}
\begin{figure}
\plotone{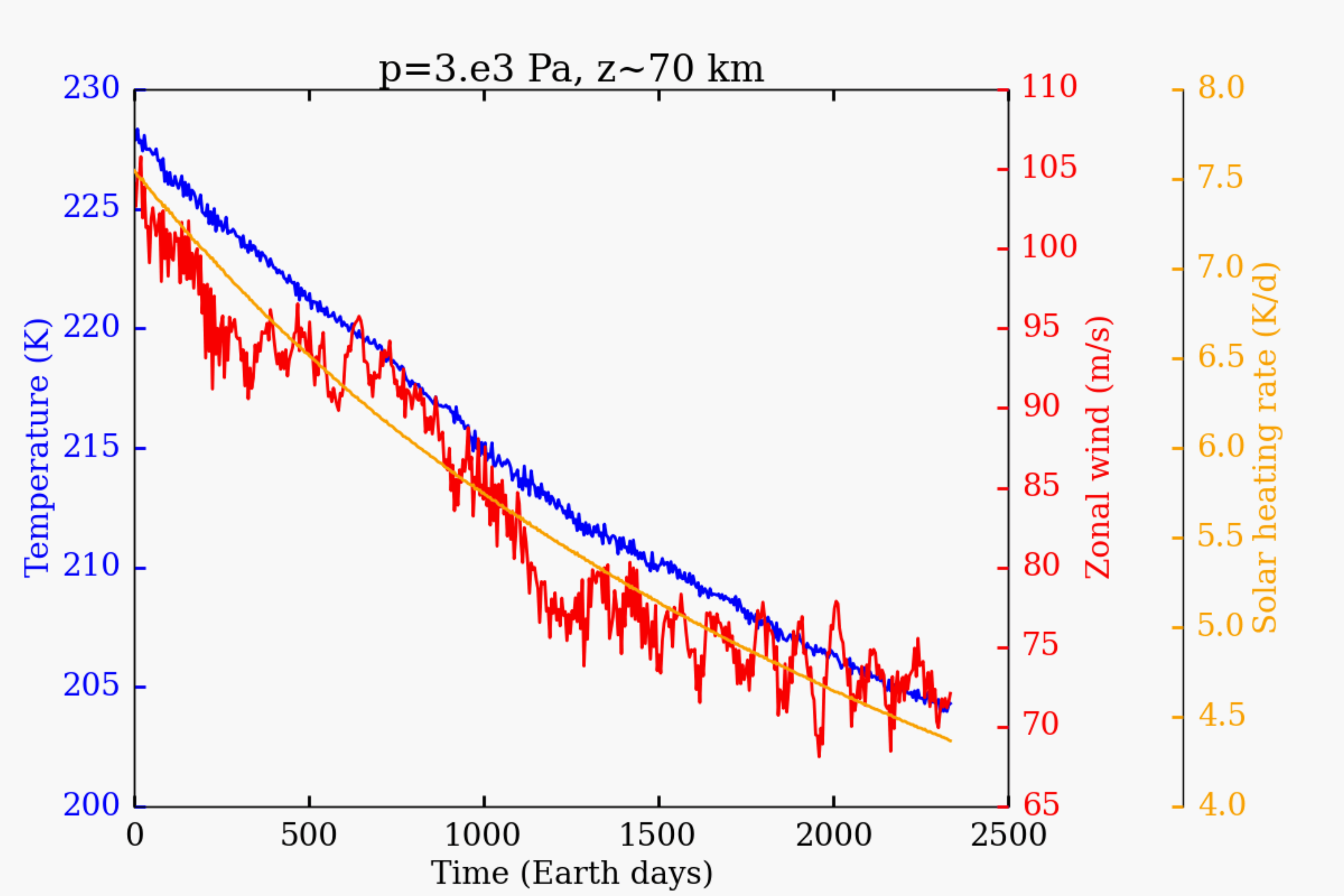}
\caption{Temporal variations of zonal mean solar heating rate (orange), wind speed (red), and temperature (blue) at 10$^{\circ}$S--20$^{\circ}$S at 30~mbar ($\sim$70~km). The solar heating rate is controlled to decrease smoothly along time by 40\% from the reference condition in the IPSL-Venus GCM (Appendix~\ref{Appendix1}). Simultaneous variations in temperature and zonal wind speed are shown together. \label{fig:zonave_T_U_dTsw}}
\end{figure}

The observed 365-nm albedo variations should \replaced{results}{result} in solar heating variations near the cloud top level as shown in Section~\ref{subsec:RTMresults}. We note that two long-term trends occurred in parallel; in the period when the 365 nm albedo \replaced{have}{had} declined, leading to increases in solar heating, the long-term cloud top zonal wind was observed to increase from 80--90~m/s to $\sim$110~m/s between 2007 and 2012 around local noon at 20$^{\circ}$S \citep{Khatuntsev13,Kouyama13,Hueso15}. Likewise, between 2014 and 2016, when the 365 nm albedo \replaced{have}{had} increased, leading to decreases in solar heating, the zonal wind speed was observed to slow down from 110 to 100~m/s \citep{Horinouchi18}. This wind speed variation is qualitatively consistent with the expected change in vertical shear associated with cyclostrophic balance. As the strongest solar heating occurs at low latitudes, the low latitudinal heating can alter the pole-to-equator gradient of temperature. The increased low latitudinal solar heating inferred from the 2011-2013 period should have increased the meridional temperature gradient, which then increases the equilibrium vertical shear, leading to increase of wind speed. Additionally there would be contributions of change of momentum flux associated with the thermal tide, which has been expected to participate in the angular momentum budget. To test these ideas, a simulation was run with the latest version of the IPSL Venus GCM \citep{GarateLopez18}. Starting from the reference simulation (see Appendix~\ref{Appendix1} for details), the solar heating rate was modified as a function of time, slowly reducing it (the whole profile at the same time, a simple test adjustment) over 20 Venusian Solar days, i.e. 2340 Earth days or 6.4 Earth years. The rate of solar heating change is $-$40\% in 6.4 years (from 7.5~K/day to 4.5~K/day in zonal mean solar heating rate), consistent with the evolution proposed in this study. The time evolution of the zonally-averaged solar heating rate in the latitudinal band between 10$^{\circ}$S and 20$^{\circ}$S and at 30~mbar ($\sim$70~km, near the cloud top level) is plotted in figure~\ref{fig:zonave_T_U_dTsw}, together with the simultaneous evolution of the temperature and zonal wind in the same region. There is a clear correlation between these variables with an amplitude only slightly higher than the observed zonal wind variations. This correlation is caused by a reduced meridional circulation that directly affects the transport of angular momentum upward and poleward, resulting in a reduction of the cloud-top zonal wind peak. Detailed analysis is described in Appendix~\ref{Appendix1}.

\subsection{Possible causes of the observed albedo variations} \label{sec:dis2_reason}
There are many intervening agents which may act in combination to produce the observed albedo variations, for example, the chemical composition and reaction rate of the unknown absorber, its interaction with or dependency on the chemical state of other atmospheric constituents, and the variability of the cloud and haze structure as a function of time. We note that the SO$_2$ abundance above the cloud top level was observed to decline from 2007 to 2012 \citep{Marcq13}, and then subsequently increase around 2016 \citep{Encrenaz19}. Since the 365~nm albedo will become relatively brighter with more abundant pure sulfuric acid aerosols which are formed through photolysis of SO$_2$, it is plausible that the observed albedo trend is linked to the long-term trend of SO$_2$ abundance.

\begin{figure}
\plotone{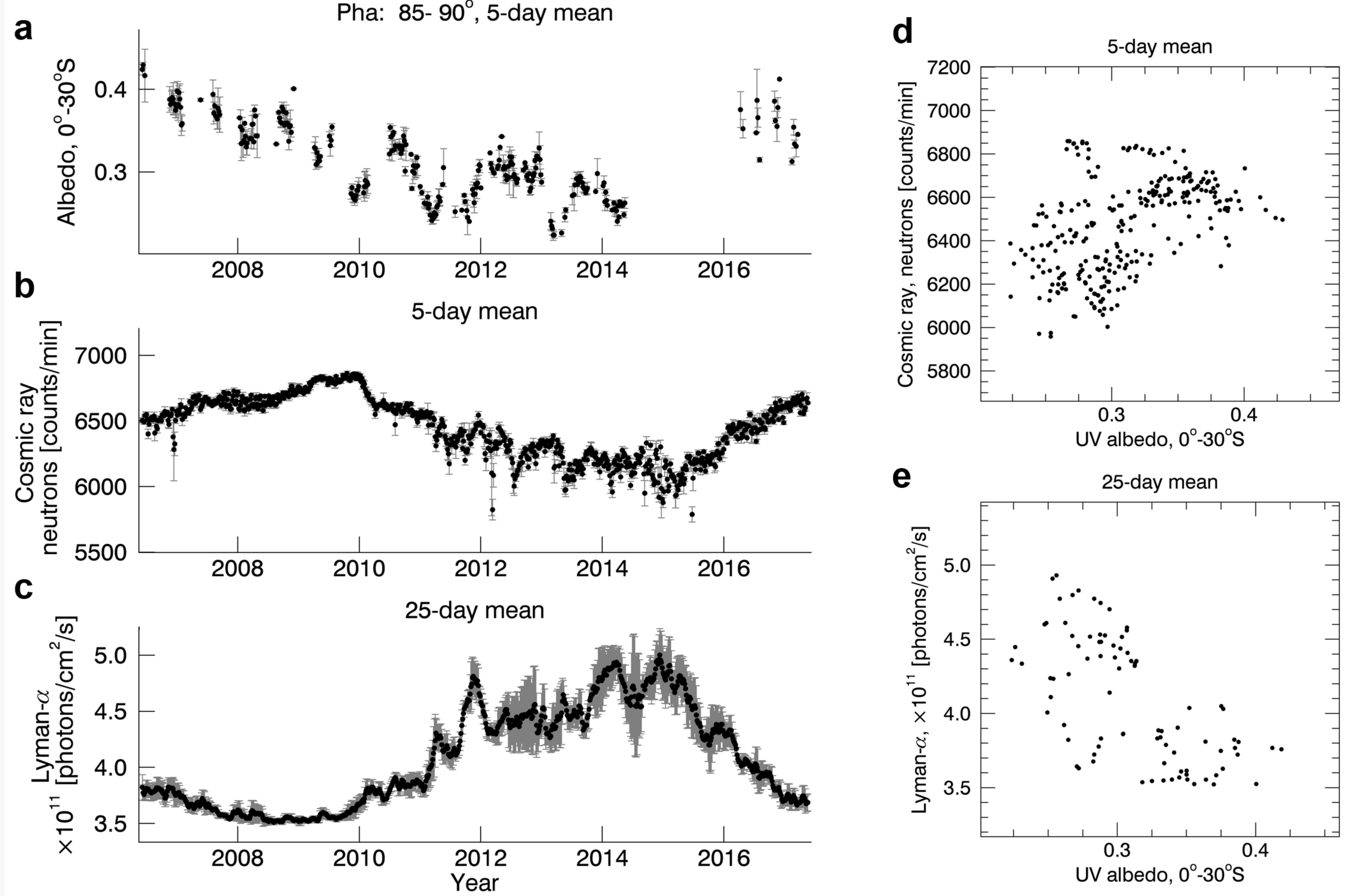}
\caption{Comparison of the 365-nm albedo $A$, neutron cosmic rays, and Lyman-$\alpha$ flux density evolution with time. On each date the 5-day mean of low latitudinal $A$ at 85--90$^{\circ}$ phase angle bin (a); the 5-day mean of cosmic ray (neutron) measured at the Oulu station (b); the 25-day mean of Lyman-$\alpha$ flux (c) is shown. A comparison between (a) and (b) is shown in (d). A comparison between (a) and (c) is shown in (e), but using the 25-day mean for consistency. See text for details. \label{fig:UV_Solar_Cosmic}} 
\end{figure}

The period of low 365~nm albedo in this study overlaps the known maximum time of solar activity cycle \citep{Jiang18} as shown in figure~\ref{fig:UV_Solar_Cosmic}. This resembles the correlation of Neptune's reflectivity and the solar activity cycle \citep{Aplin16}. Since solar EUV radiation might affect photochemical reactions involving SO$_2$ that are necessary for aerosol formation on Venus \citep{Mills07, Parkinson15}, further study is required to explore influences of solar activity cycle on the Venusian atmosphere. It is also possible that production rate of sulfuric acid aerosols is altered by galactic cosmic rays via ion-induced nucleation. Electrostatic interactions between ionized acid molecules can enhance new aerosol formations by reducing critical size and by increasing collision possibility \citep{Lovejoy04,Kirkby07}, and there are observations on H$_2$SO$_4$-H$_2$O ultrafine aerosols of less than 9~nm in diameter in Earth’s upper troposphere and stratosphere that were explained by the ion-induced nucleation \citep{Lee03}. The peak of ion production rate in the Venusian atmosphere due to galactic cosmic rays is predicted \replaced{as}{at} 62.5~km \citep{Nordheim15} with the 46--58 ion pairs cm$^{-3}$ s$^{-1}$ range of variations between solar minimum to maximum. So the upper haze aerosol formation might be effectively triggered by such ion-induced nucleation, and vary following the solar activities. Figure~\ref{fig:UV_Solar_Cosmic} shows comparisons of the 365-nm albedo at low latitudes, neutron cosmic ray detected from the Oulu station\footnote{\url{http://cosmicrays.oulu.fi}}, and Lyman-$\alpha$ flux at the Earth location\footnote{\url{http://lasp.colorado.edu/lisird/data/composite_lyman_alpha/}}. A 5-day mean is applied to compare the 365-nm albedo and cosmic ray to compensate the atmospheric rotation rate on Venus. The solar rotation rate (25-day) mean is applied for the comparison of the 365-nm albedo and Lyman-$\alpha$ flux, to take into account the different locations of Venus and of the Earth respect to the Sun. The results shows that the 365-nm albedo $A$ has a negative correlation with Lyman-$\alpha$ flux, and a positive correlation with neutron cosmic rays, but these may act together with the mesospheric SO$_2$ gas influences.

\subsection{Comparison with other planets} \label{sec:dis3_comparisonwithotherplanets}
Regardless of the cause of the observed albedo changes, the range of albedo variation on Venus is surprising. On the Earth, clouds play a considerable role as a buffer of possible climate variations and are also a regulator of the solar energy distribution \citep{Stephens14}. However, the clouds on Venus are different; rather than supporting a stable solar heating rate, drastic variations of solar heating seem to occur as inferred from the 365-nm albedo. The astounding nature of the albedo variation results we present here is further emphasized by results derived from other planetary albedo studies in the Solar System where weaker long-term albedo variations were observed. For example, at Neptune the observed magnitude varied by $\pm$0.02 (corresponding to $\pm$2\% changes in flux) at the blue (472$\pm$10~nm) and red (551$\pm$10~nm) filters over 1972--2014 \citep{Aplin16}, and at Mars where the surface albedo varied by 10\% at the red filters (575–-675 and 550--700 nm) from 1976--1980 to 1999--2003 \citep{Geissler05}.

\subsection{Further studies} \label{sec:dis4_furtherstudies}
In addition to the impact of the solar heating rates at the cloud top level, the vertical profile of solar flux on Venus, down to the surface, should also be altered by the observed cloud top albedo changes. Such solar flux variations may explain unbalanced net radiative energy below the clouds \citep{Lee17}, and therefore further investigation is needed to understand the true impact of the albedo changes on the entire lower atmosphere of Venus.

Our study focuses on the observed 365-nm albedo and its direct impacts on solar heating at the equator. This does not cover detailed modeling of net radiative forcing, such as cooling rate changes \citep{Haus17}, that would impact on microphysical and photochemical processes. Such studies must be completed to accurately infer the true impact of the solar heating on cloud formation and climate. The work we present here provides a foundation for future in-depth studies of links between Venus’ 365 nm albedo and processes that directly \edit2{impact\deleted{impacting}} Venus’ climate.

\section{Summary} \label{sec:summary}
We present the intense decadal variation of Venus' 365-nm albedo between 2006 and 2017; the maximum albedo occurred in 2006--2007, the minimum in 2011--2014, and the recovery of albedo in 2016--2017 to the level in 2008--2009. This trend is consistent among four independent UV instruments; VMC and MASCS in 2007, and VMC, STIS, and UVI in 2011, either using disk-resolved or disk-integrated data. We discard the previously suggested sensitivity degradation of VMC \citep{Shalygina15}, and propose a new calibration correction factor for VMC in this study. The ranges of albedo variations are $\sim$0.2--0.4 at low latitudes, and $\sim$0.3--0.6 at high latitudes in 2006--2017, so albedo had been varied by a factor of two over the last decade. The whole-disk albedo also shows a similar trend, changed from \edit2{$-$}30\% to +20\% compared to the mean value in 2016--2017, meaning that the albedo variation occurred \edit2{on\deleted{in}} a global scale.

Our one dimensional line-by-line radiative transfer model calculations \edit2{reveal\deleted{reveals}} this level of albedo variation can alter solar heating rate from \edit2{$-$}25\% to +40\% due to the broad absorption spectrum of the unknown absorber from UV to visible range. We suggest that this solar heating rate \edit2{variation\deleted{variations}} can be a cause of the observed long-term zonal wind speed variation at low latitudes that increased from 80--90~m/s in 2007 to $\sim$110~m/s in 2012, and then decreased to 100~m/s in 2016--2017. Wind speed increased during the low albedo time, when solar heating was stronger than average, implying that increased solar heating may play a role in wind speed changes through cyclostrophic balance, enhanced thermal tide, or vertical momentum transport. We show the results of Venus GCM simulations, which support the linear relationship between solar heating rate and zonal wind speed.

The observed 365-nm albedo variations might be caused by variations of SO$_2$ gas abundance above the clouds. We also suggest links between the 365~nm albedo, and the Solar Cycle and consequent galactic cosmic ray density variations. Continuous 365~nm observations are necessary to clarify the mechanism of the 365~nm albedo variations.

\appendix
\section{Brief description on IPSL Venus GCM, and analysis of the simulation results.}
\label{Appendix1}
Institut Pierre Simon Laplace (IPSL) Venus Global Climate Model (GCM) has successfully demonstrated the development of strong zonal wind near the cloud top level \citep{Lebonnois16}. Upgrading the GCM, \citep{GarateLopez18} employed latitudinally varying cloud structures \citep{Haus14}. The authors also used solar heating look-up table \citep{Haus15b} according to the latitudinal cloud structures. Thermal cooling is based on the net-exchange rate (NER) formalism \citep{Eymet09} with additional continua. This last version of the IPSL Venus GCM was able to simulate prominent cold band surrounding the poles of Venus close to observations \citep{GarateLopez18}.

\begin{figure}
\plotone{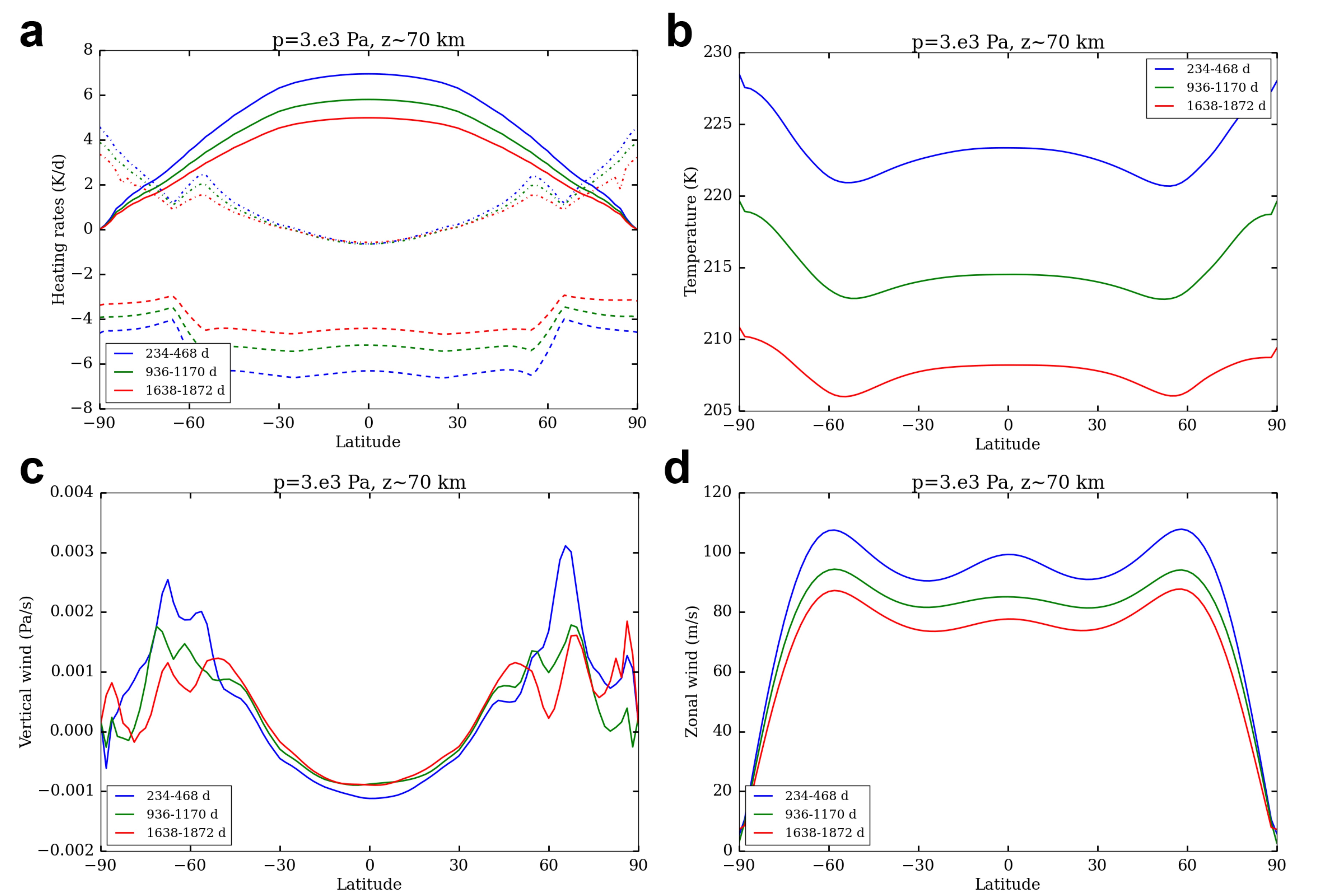}
\caption{Latitudinal heating rate (a), temperature (b), vertical winds (c), and zonal winds (d) at 30~mbar ($\sim$70~km) in the same simulation shown in figure~\ref{fig:zonave_T_U_dTsw}. All parameters are averaged zonally over 2~Venusian solar days as shown in the legend of each panel. In (a), solar heating (solid line), thermal heating (dashed line), and adiabatic heating (dotted line) are compared together, and negative values mean cooling. \label{fig:GCM_result4}}
\end{figure}

In this latest IPSL Venus GCM, only solar heating has been reduced, \replaced{mimicing}{mimicking} the expected solar heating rate variations shown in this manuscript. Figure~\ref{fig:zonave_T_U_dTsw} presents the linear correlation among solar heating, temperature, and zonal wind speed. To interpret these correlations, the time variations of the latitudinal profiles of zonally and temporally (over 2~Venusian solar days) averaged heating rates (solar, infrared and dynamical terms of the energy budget), temperature, and vertical and zonal winds are plotted in figure~\ref{fig:GCM_result4} at 30~mbar.

Looking at the different heating rates (figure~\ref{fig:GCM_result4}a), it appears that in average, the decrease of the solar heating is mostly compensated by a decrease of the infrared cooling, corresponding to the decrease in temperature (figure~\ref{fig:GCM_result4}b). At mid- to high-latitude, though, the dynamical term associated with averaged meridional and vertical motions is not negligible, and therefore the decrease of the solar heating is compensated by an impact on the averaged meridional and vertical winds. A reduction of the amplitude of the vertical wind is seen in figure~\ref{fig:GCM_result4}c, except between 30 and 50$^\circ$ of latitude, which may indicate some impact here of changes in the meridional energy budget. This reduction of the mean meridional circulation has a direct impact on the transport of angular momentum upward and poleward, inducing a reduction of the cloud-top zonal wind peak (figure~\ref{fig:GCM_result4}d). Regardless \edit2{of\deleted{on}} the simplicity of the simulation set-up, the results explain how solar heating variations can affect zonal winds. It will be important to compare with long-term temperature trend analysis that may be available in near future using night side temperature field retrieved from VIRTIS-H/Venus Express \citep{Migliorini12} or radio occultation temperature profiles from VeRa/Venus Express \citep{Tellmann12}.

\acknowledgments
Y.J.L thanks T. Satoh, R. Lorenz, N. Ignatiev, R. Hueso, and C.C.C. Tsang for helpful comments. K.-L. J. was supported by NASA/PS program, grant NNX16AK82G, for the analysis of the 365~nm STIS data. S.P.-H. was supported by the Spanish MICIIN project AYA2015-65041-P (MINECO\/FEDER, 695UE), and Grupos Gobierno Vasco IT-765-13. S.Le. acknowledges support from INSU-PNP, IPSL VGCM computations were done thanks to the High-Performance Computing (HPC) resources of Centre Informatique National de l'Enseignement Sup\'erieur (CINES) under the allocation n$^{\circ}$A0040110391 made by Grand \'Equipement National de Calcul Intensif (GENCI). J.P. \replaced{aknowledges}{acknowledges} the JAXA's International Top Young Fellowship. Solar heating calculations were performed on a Supermicro SuperServer Intel(R) Xeon(R) CPU E5-2620 v4 funded through JAXA's International Top Young Fellowship. T. H., M. T., and T. I. acknowledge JSPS grant 16H02231. S.Li. was supported by a NASA Participating Scientist in Residence Grant (NNX16AC79G). E.M acknowledges support from CNES and INSU-PNP for SPICAV data analysis.

\bibliography{paper}
\end{document}